\documentclass[12pt,twoside,a4paper]{article}
\newcommand{\ber}{\begin{eqnarray}}
\newcommand{\eer}{\end{eqnarray}}

\newcommand{\be}{\begin{equation}}
\newcommand{\ee}{\end{equation}}

\newcommand{\eps}{\epsilon}

\newcommand{\cs}{{\bf CS }}
\newcommand{\ppt}{{DP }}
\newcommand{\bm}[1]{\mbox{\boldmath{$#1$}}}

\usepackage{amsmath,amssymb,amsfonts,epsfig}
\usepackage{verbatim}
\usepackage{graphicx}
\usepackage{graphics}
\usepackage{url}
\usepackage[sort&compress]{natbib}
\usepackage[footnotesize,bf]{caption}

\title{Rare-Allele Detection Using Compressed Se(que)nsing}

\author{
 
 Noam Shental\footnotemark[1] \quad Amnon Amir\footnotemark[2] \quad
 Or Zuk\footnotemark[3]}

\begin{document}
\maketitle \footnotetext[1]{Department of Computer Science,
The Open University of Israel, Raanana, 43107, Israel\newline  shental@openu.ac.il}
\footnotetext[2]{Department of Physics of Complex Systems,
Weizmann Institute of Science, Rehovot, 76100, Israel\newline  amnon.amir@weizmann.ac.il}
\footnotetext[3]{Broad Institute of MIT and Harvard, Cambridge, 02142, MA, USA\newline  orzuk@broadinstitute.org}

\abstract
Detection of rare variants by resequencing is important for the identification of individuals
carrying disease variants.
Rapid sequencing by new technologies enables low-cost resequencing of target regions,
although it is still prohibitive to test more than a few individuals.
In order to improve cost trade-offs, it has recently been suggested to
apply pooling designs which enable the detection of carriers of rare alleles in groups of individuals.
However, this was shown to hold only for a relatively low number of individuals in a pool,
and requires the design of pooling schemes for particular cases.

We propose a novel pooling design, based on a {\it compressed sensing} approach,
which is both general, simple and efficient.
We model the experimental procedure and show via computer simulations
that it enables the recovery of rare allele carriers out of larger groups
than were possible before, especially in situations where high coverage is obtained 
for each individual. 

Our approach can also be combined with barcoding techniques to enhance
performance and provide a feasible solution based on current resequencing costs.
For example, when targeting a small enough genomic region ($\sim\!\! 100$ base-pairs) and using only 
$\sim\!10$ sequencing lanes and $\sim\!\!10$ distinct barcodes,
one can recover the identity of $4$ rare allele carriers out of a population of over $4000$ individuals.

\section{Introduction}
\label{sec:intro}
Genome-Wide Association Studies (GWAS)~\cite{hirschhorn2005genome} have been
successfully used in recent years to detect associations between genotype and phenotype,
and numerous new alleles have been found to be linked to various human
traits~\cite{klein2005complement,burton2007genome,sladek2007genome}.
However, genotyping technologies are limited only to those variants
that are predetermined and prioritized for typing,
which results in a bias towards typing of {\it common} alleles.

Although many common alleles were lately found to have statistically significant
associations with different human traits,
they were thus far shown to explain only a small fraction of most traits' heritability content.
This, together with other theoretical and empirical arguments, raise the possibility that in fact
{\it rare} alleles may play a significant role in the susceptibility of human individuals to many common diseases
~\cite{cohen2004multiple,mcclellan2007schizophrenia,bodmer2008common,li2009discovery}.
Discovering and genotyping of rare alleles may therefore be of great bio-medical interest,
however such studies require genotyping of large human populations - a task considered infeasible until recently.

This state of affairs may change dramatically as we are currently witnessing a rapid revolution
in genome sequencing due to emerging new technologies.
Sequencing throughput at a given cost is growing at an exponential rate,
in similar to Moore's law for computer hardware \cite{mardis2008impact}.
At present next-generation sequencing technologies ~\cite{margulies2005genome,gunderson2004decoding,harris2008single}
utilize massively parallel reading of short genomic fragments to achieve
several orders of magnitude higher throughput at the same cost as previous Sanger sequencing machines~\cite{mardis2008impact}.
The availability of cheap, high-throughput rapid sequencing methods leads
to a change in the way researchers  approach various biological problems,
as it enables addressing questions which were infeasible to be studied before.

Next-generation sequencing opens the possibility to obtain the genomic sequences
of multiple individuals along specific regions of interest.
This approach, often called resequencing, is likely to provide an extensive amount of novel information
on human genetic variation. In particular, the ability to resequence a large number of individuals will enable
the study of {\it rare} alleles in human populations. Resequencing of large populations can thus fill
a gap in our knowledge by allowing us to discover and type these rare variants,
often with frequencies well below $1\%$, at given predefined regions. 
Of particular interest are regions around loci that have previously been established  for involvement in disease, 
as they can be resequenced across a large population to seek novel variations.

Another important application of interest is resequencing a set of specific {\it known}
Single-Nucleotide-Polymorphisms (SNPs) with low minor allele frequency which
are known or suspected to be important for a certain trait. In this case we are interested in
identifying individuals carrying these alleles out of a very large group of individuals.
For example, this may assist in scanning of large populations for individuals carrying
 certain risk alleles for a potentially lethal disease.
For the sake of clarity in this paper we focus on this application,
although discovery and typing of unknown rare variants can also be applied following the same lines.

Current next-generation sequencing technologies provide throughput on the order of millions of reads
in a single `run' or `lane'~\cite{mardis2008impact},
where a sequence read is typically a short consecutive DNA fragment of
 a few dozens to a few hundreds nucleotides.
In addition, novel experimental procedures enable targeted selection
of pre-defined genomic regions prior to sequencing~\cite{albert2007direct,gnirke2009solution,Ng:01}.
These methods, also called `hybrid capture' or `hybrid selection',
enrich significantly the DNA or RNA within the regions of interest
and minimize the number of reads `wasted' on fragments residing outside these regions.
Together, these high-throughput technologies have made the identification of
carriers over a pre-defined region a feasible, yet still an expensive task.

A naive but costly option is to utilize one lane per individual.
However, when considering a population of hundreds or thousands of individuals,
such an approach is prohibitively expensive.
Moreover, since resequencing is typically performed on targeted regions rather
than the whole genome, throughput requirements to sequence
an individual are much lower than the capacity of a single lane,
thus the naive approach is also highly inefficient.

In such cases, ``pooled'' sequence runs may offer a more feasible approach.
In ``pooled DNA'' experiments, DNA from several
individuals is mixed and sequenced together on a single sequencing lane.
Pooled genotyping has been used to
quantify previously identified variations and study allele frequency
distributions in populations~\cite{norton2002universal,yang2007association,shaw1998allele}.
Given a measurement for each allele, it is possible to estimate the average frequency
of the allele in those individuals participating in the pool.
However, traditional pooled sequencing was used only to infer
the {\it frequency} of rare alleles in a population,
and did not give means to recover the {\it identity} of rare allele carriers. In this work we
focus on the latter task, of {\it identifying} rare allele carriers by sequencing pooled DNA.

The field of {\it group testing}~\cite{du2000combinatorial} aims to tackle
this problem of identifying individuals carrying a certain trait out of a group,
by designing an efficient set of test, i.e., pools.
This field which dates back to the mid 20'th century has
applications in biology~\cite{du2000combinatorial}, streaming algorithms~\cite{muthukrishnan2005data}
and communications (See \cite{Gilbert:01} for a comprehensive survey.)
Recently, several works have tried to use resequencing-based group testing methods in order to identify rare allele carriers.

Prabhu and Pe'er~\cite{Prabhu:01} offered to use overlapping pools, elegantly designed
based on Error-Correcting-Codes, to enable the recovery of
a single rare-allele carrier from multiple pools. Individuals are represented in multiple pools,
where the composition of different pools is constructed in a way which provides a unique pooling `signature'
for each individual.
This carefully designed scheme enables the recovery of a rare allele carrier by
observing the presence of reads containing the rare allele in these `signature' pools. Their design offers a
significant saving in resources, as it enables the recovery of a single carrier out of $N$ individuals, by
using only $O(\log N)$ pools. It is, however, limited to the case of a single rare allele carrier within the
group, and the problem of detecting multiple (albeit few) carriers remained unsolved.

In another approach by Erlich et al.~\cite{Erlich:01} a clever barcoding scheme
combined with pooling was used,
in order to enable the identification of each sample's genotypes.
When using barcoding, each sample is ``marked'' by a unique short sequence identifier, i.e., barcode,
 thus upon sequencing one can identify the origin of each read according
to its barcode, even when multiple samples are mixed in a single lane. Ideally, one could assign a different
barcode to each individual sample, and then mix many samples in each lane while keeping the identity of each
read based on its barcode. However, barcoding is a costly and laborious procedure, and one wishes
to minimize the number of barcodes used.
It was therefore suggested in \cite{Erlich:01} to barcode different
pools of samples (rather than individual samples), thus allowing the barcode to identify
the pool from which a certain read was obtained, but not the identity of the specific sample.
Efficient algorithms based on the Chinese-Remainder-Theorem enable the accurate recovery of rare
allele carriers, where both the total number of pools and the number of individual samples participating
in each pool were kept low - the identification of $N$ individual genotypes was obtained by using
$O(\sqrt{N})$ different pools with $O(\sqrt{N})$ individuals per pool.

In this work we present a different approach to recovering the identity of individuals
carrying rare variants, based on Compressed Sensing (\cs\!\!).
\cs and group testing are intimately connected~\cite{Gilbert:02},
yet this approach was not studied in the context of rare allele identification.
Our work extends the idea of recovering the identity of rare-allele carriers using
overlapping pools beyond the single carrier case analyzed in \cite{Prabhu:01},
and deals with heterozygous or homozygous rare alleles.
The \cs pooling approach enables testing of a larger cohort
of individuals, thus identifying carriers of rarer SNPs.
The \cs paradigm also adapts naturally and efficiently to the addition of barcodes.
We treat barcodes as splitting a given lane into many different lanes of
approximately equal sizes in terms of number of reads - thus barcoding effectively
boosts our number of lanes.

Compressed Sensing \cite{Candes:01,Donoho:01} is a new emerging and very active field of research,
with foundations in statistics and optimization.
New developments, updates and research papers in \cs appear literally on a daily basis, in
various websites (e.g. \url{http://dsp.rice.edu/cs}) and blogs (e.g. \url{http://nuit-blanche.blogspot.com/}).
Applications of \cs theory can be found in many distantly related fields such as magnetic
resonance imaging \cite{Lustig:01}, single pixel cameras \cite{Duarte:01}, geophysics \cite{Lin:02},
astronomy \cite{Bobin:01} and multiplexed DNA microarrays \cite{Dai:01}.

In \cs one wishes to efficiently  reconstruct an unknown vector of values
$\textbf{x}=(x_1,...,x_N)$, assuming that $\textbf{x}$ is {\it sparse}, i.e., have at most $s$ non-zero entries,
for some $s\ll N$. It has been shown that $\textbf{x}$ can be reconstructed using $k\ll N$ basic operations termed
``measurements'', where a measurement is simply the output $y$ of the dot-product of the (unknown vector) $\textbf{x}$ with
a known measurement vector $\textbf{m}$, $y = \textbf{m} \cdot \textbf{x}$.
By using the values of these $k$ measurements and their corresponding $\textbf{m}$'s, it is then possible
to reconstruct the original sparse vector $\textbf{x}$.

Mapping of group testing into a \cs setting is simple.
The entries of $\textbf{x}$ contain the genotype of each individual at a specific genetic locus
and are non-zero only for minor allele carriers, thus since we are interested in rare alleles
$\textbf{x}$ is indeed sparse. A measurement in our setting corresponds to sequencing the DNA of
a pool of several individuals taken together, hence the measurement vector represents the individuals
 participating in a given pool and the output of the measurement is proportional
to the total number of rare alleles in the pool.
Our basic unit of operation is a single `run' or `lane', which is used to sequence $L$ pre-defined
different loci in the genome, whether consecutive in one specific region, taken from different regions
or surrounding different SNPs.
We treat each of the $L$ different loci separately and reconstruct $L$ different
vectors $\textbf{x}$, thus the amount of computation increases linearly with the number of loci of interest.

Formulating the problem in terms of \cs opens the door to utilizing
this fascinating and seemingly almost magical theory for our purposes.
In particular, from a theoretical perspective one can use \cs bounds to estimate
the number of samples and lanes needed for reconstruction, and the robustness of the
reconstruction to noise.
From a more practical point of view, we can apply numerous algorithms and techniques available
for \cs problems, and benefit from the development of faster and more accurate reconstruction algorithms
as the state of the art is constantly improving (see e.g. \cite{Becker:01}.)
We believe that \cs is a suitable approach for identifying rare allele carriers,
and hope that this paper is merely a first step in this direction.

In this work we present results of extensive simulations which aim to explore the benefits and limitations of
applying \cs for the problem of identifying carriers of rare alleles in different scenarios.
We provide a detailed model of the experimental procedure typical to next generation sequencing
and find scenarios in which the benefit of applying \cs is overwhelmingly large (up to over $\sim\! 70X$ improvement)
compared to the naive one-individual-per-lane approach.
We also show that our method can be used in addition to barcodes,
and provide a significant improvement over applying each of these methods by itself.

The paper is organized as follows:
Section~\ref{sec:methods} presents \cs in the context of identifying carriers
of rare alleles, and the specific details of our proposed pooling design, genotype
reconstruction algorithm and the noise model reflecting the pooled sequencing process.
Simulation results are presented in Section~\ref{sec:Results}, and provide evidence for the efficiency
of our approach along a wide range of parameters.
Finally, Section~\ref{sec:discussion} offers conclusions and outlines possible directions for future research.

\section{Methods}
\label{sec:methods}
We first provide a short overview of \cs\!\!,
followed by a description of its application to our problem of identifying rare alleles,
and the corresponding mathematical formulation including a noise model reflecting the sequencing process.
Finally we show how one performs reconstruction while utilizing barcoding.

\subsection{The Compressed Sensing Problem}
\label{sec:csIntro}
In a standard \cs problem one wishes to reconstruct a sparse vector $\textbf{x}$ of length $N$,
by taking $k$ different measurements $y_i = \textbf{m}_{i} \cdot \textbf{x}, \: i=1,\ldots,k$.
This may be represented as solving the following set of linear equations
\be
M \bf{x} = \textbf{y}
\label{eq:cs_classic}
\ee

where $M$ is a $k \times N$ {\it measurement matrix} or {\it sensing matrix}, whose rows are the different $\textbf{m}_i$'s
(as a general rule, we use upper-case letters to denote matrices: $M,E,..$, lower boldface letters to denote vectors:
$\textbf{x}, \textbf{y}, ..$ and lower case to denote scalars: $x,y,x_i$.)

Typically in \cs problems, one wishes to reconstruct $\textbf{x}$ from a small
number of measurements, i.e. $k\ll N$, hence the linear system (\ref{eq:cs_classic}) is under-determined,
namely there are `too few' equations or measurements and $\textbf{x}$ cannot be recovered uniquely.
However, it has been shown that if $\textbf{x}$ is sparse,
and $M$ has certain properties, the original vector $\textbf{x}$ can be recovered uniquely
from Eq.~(\ref{eq:cs_classic}) \cite{Candes:01,Donoho:01}.
More specifically, a unique solution is found in case $k > C s \log (N/s)$,
where $C$ is a constant, and $s$ is the is the number of non-zero entries in $\textbf{x}$.
This somewhat surprising result stems from the fact that the desired solution $\textbf{x}$ is sparse,
 thus contains less `information' than a general solution. Therefore,
one can `compress' the amount of measurements or ``sensing'' operations required for
the reconstruction of $\textbf{x}$.

A sensing matrix which  allows for a correct reconstruction of $\textbf{x}$
posses a property known as ``Uniform Uncertainty Principle'' (UUP) or Restricted Isometry Property (RIP)~\cite{candes2005decoding,candes2005stable}.
Briefly, UUP states that any subset of the columns of $M$ of size $2s$ form a matrix which
is almost orthogonal (although since $k<N$ the columns cannot be perfectly orthogonal),
which, in practice, makes the matrix $M$ ``invertible'' for sparse vectors $\textbf{x}$.
The construction of a ``good'' sensing matrix is an easy task when one is able to use randomness.
As an example of a UUP matrix consider a Bernoulli matrix, namely a matrix whose entries
are independent random variables set to be $1$ or $-1$ with probability
$0.5$\footnote{It is common to require orthonormality of the columns therefore the entries
should also be divided by $\sqrt{k}$.}.
It is known that almost any instance of such a random matrix will satisfy UUP
with an overwhelming probability~\cite{donoho2006most,candes2006near} (the same is true when each entry
in the matrix is a standard Gaussian random variable.)

Once $M$ and $\textbf{y}$ are given,
\cs aims to find the sparsest possible $\textbf{x}$ which obeys Eq.~(\ref{eq:cs_classic}).
This can be written as the following optimization problem:
\be
{\textbf{x}}^* = \underset{\textbf{x}}{argmin} ||\textbf{x}||_0 \quad s.t. \: \: M \textbf{x}=\textbf{y}
\label{eq:cs_problem}
\ee

where the $\ell_0$ norm $||\textbf{x}||_0 \equiv \sum_{i} 1_{\{x_i \neq 0\}}$ simply counts the
number of non-zero elements in $\textbf{x}$.

Problem (\ref{eq:cs_problem}) involves a non-convex $\ell_0$ term and can be shown to be
computationally intractable in general \cite{candes2005error}. However, another impressive breakthrough
of \cs theory is that one can relax this constraint to the closest convex $\ell_p$ norm, namely the
$\ell_1$ norm, and still get a solution which, under certain conditions, is identical to the solution of
problem (\ref{eq:cs_problem}). Hence the problem is reformulated as the following $\ell_1$ minimization problem,
which can be efficiently solved by convex optimization techniques:
\be
{\textbf{x}}^* = \underset{\textbf{x}}{argmin} ||\textbf{x}||_1 \quad s.t. \: \: M \textbf{x}=\textbf{y}
\label{eq:cs_problem_L1}
\ee

In most realistic \cs problems measurements are corrupted by noise, hence Eq.~(\ref{eq:cs_classic})
is replaced by $M \textbf{x} + \bm{\eta} = \textbf{y}$, where $\bm{\eta}=(\eta_{1},\ldots,\eta_{k})$
are the unknown errors in each of the $k$ measurements, and the total measurement noise, given by
the $\ell_2$ norm of $\bm{\eta}$ is assumed to be small.
Therefore, the optimization problem is reformulated as follows:
\be
{\textbf{x}}^* = \underset{\textbf{x}}{argmin} ||\textbf{x}||_1 \quad s.t. \: \: ||M \textbf{x} - \textbf{y}||_2 \leq \eps
\label{eq:cs_problemFinal}
\ee
where $\eps > 0$ is set to be the maximal level of noise we are able to tolerate, while still obtaining a sparse solution.
It is known that \cs reconstruction is robust to noise,
 thus adding the noise term $\eps$ does not cause a breakdown of the \cs machinery,
but merely leads to a possible increase in the number of measurements $k$ \cite{candes2005stable}.

Many efficient algorithms are available for problem~(\ref{eq:cs_problemFinal}) and enable a practical solution
even for large matrices, with up to tens of thousands of rows. We have chosen to work with the commonly used
GPSR algorithm \cite{GPSR:01}.

\subsection{Rare-Allele Identification in a \cs Framework}
\label{sec:rare_cs_framework}
We wish to reconstruct the genotypes of $N$ individuals at a specific locus.
The genotypes are represented by a vector $\textbf{x}$ of length $N$,
where $x_i$ represents the genotype of the {\it i}'th individual. We denote the reference allele
by $A$, and the alternative allele by $B$. The possible entries of $x_{i}$ are $0$, $1$ and $2$,
representing a homozygous reference allele ($AA$), a heterozygous allele ($AB$)
and a homozygous alternative allele ($BB$), respectively.
Hence, $x_i$ counts the number of (alternative) $B$ alleles of the {\it i}'th individual,
and since we are interested in rare minor alleles, most entries $x_{i}$ are zero.
In classic \cs the unknown variables are typically real numbers.
The restriction on $\textbf{x}$ in this case is expected to reduce the number of measurements
needed for reconstruction and may
also enable using faster reconstruction algorithms, as it is known
that even a weaker restriction, namely that all entries are positive,
already simplifies the reconstruction problem \cite{Donoho:positive}.

The sensing matrix $M$ is built of $k$ different measurements represented by the rows
of $M$. The entry $m_{ij}$ is set to $1$ if the {\it j}'th individual participates in {\it i}'th measurement,
and zero otherwise. Each measurement includes a random subset of individuals, where the probability to include
a certain individual is $0.5$. Hence, $M$ is equivalent to the Bernoulli matrix mentioned in Section~\ref{sec:csIntro},
which is known to be a ``good'' sensing matrix\footnote{This is easily seen by using the simple linear
transformation $x \to (2x-1)/\sqrt{k}$.} (another type of a sensing matrix, in which only $\sim\!\! \sqrt{N}$ elements
in each measurement are non-zero is considered in Section~\ref{sec:Results}).

In practice, measurements are performed by taking equal amounts of DNA
from the individuals chosen to participate in the specific pool,
thus their contribution to the mixture is approximately equal.
Then, the mixture is amplified using PCR, which ensures that the amplification
bias generated by the PCR process affects all individuals equally \cite{Ingman:01}.
Finally, DNA of each pool is sequenced in a separate lane,
and reads are mapped back to the reference genome (this may be performed using
standard alignment algorithms such as MAQ \cite{li2008mapping}.)
For each locus of interest we record the number of reads containing the rare allele
together with the {\it total} number of reads covering this locus in each pool, denoted by $r$.
These numbers provide the measurement vector $\textbf{y}$ representing the $k$ frequencies obtained for
this locus in the $k$ different pools.
The measurement process introduces various types of noise which we model in the next section.

\begin{figure}[thb!]
\begin{center}
\psfig{file=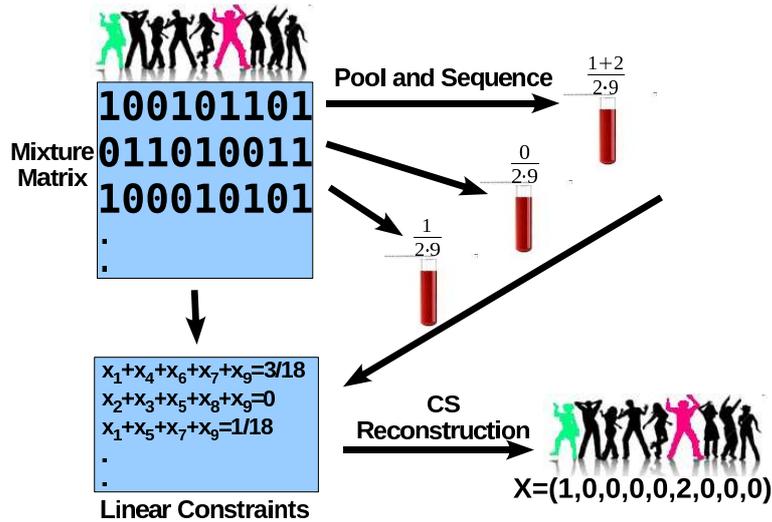,width=10cm}
\caption{Schematics of the \cs based pooling method.
Consider the case of $9$ people,
out of which one is a heterozygotic carrier of the rare SNP (marked green),
and another one who is a homozygous alternative allele carrier (marked red.)
Each sample is randomly assigned to a pool with probability $0.5$, as described by the sensing matrix.
For example, individuals $1,4,6,7$ and $9$ are assigned to the first pool.
The DNA of the individuals participating in each pool is mixed,
and the fraction of rare alleles in each pool is measured.
For example, the first pool contains the two carriers,
hence the frequency of the $B's$ is $1+2$ out of the $2\times 9$ alleles.
The sensing matrix and the resulting frequencies are incorporated into an underdetermined set of
linear constraints, from which the original rare SNP carriers are reconstructed.
\label{fig:pool_scheme}}
\end{center}
\end{figure}

For each locus, our goal is to reconstruct the vector $\textbf{x}$,
given the sensing matrix $M$ and the measurement vector $\textbf{y}$,
while realizing that some measurement error $\eps$ is present (see Eq.~(\ref{eq:cs_problemFinal}).)
Our experimental design is illustrated in Fig.~\ref{fig:pool_scheme},
and the following section describes its mathematical formulation.

\subsection{Mathematical Formulation of Our Model}
\label{sec:rare_cs_model}
The model presented here, including the range of parameters chosen, aims at reflecting the sequencing
process by the Illumina technology \cite{gunderson2004decoding}, but may also be applied to other
next generation technologies as well. It is similar, but not identical, to the model presented in \cite{Prabhu:01}.
For clarity of presentation, we first describe our model while ignoring the different
experimental noise factors, and these are added once the model is established.

Let $\textbf{x}$ be the unknown sparse genotypes vector,
as described in Section~\ref{sec:rare_cs_framework}.
The fraction of individuals with the rare allele is denoted $f$, thus the
vector $\textbf{x}$ has $s = f N$ non-zero elements\footnote{Here and throughout the paper we define the fraction
of {\it individuals} (rather than alleles) as the `rare-allele-frequency' - thus the fraction of alternative
{\it alleles}, assuming Hardy-Weinberg (HW) equilibrium, is in fact $1-\sqrt{1-f}$.}.
$M$ is a $k \times N$ Bernoulli sensing matrix,
and denote by $\hat{M}$ the normalized version of $M$ whose entries
represent the fraction of each individual's DNA in each pool:
\be
\hat{m}_{ij} \equiv \frac{m_{ij}}{\sum_{j=1}^{N} m_{ij}}
\ee

Assume that the mixing of DNA is perfect and unbiased,
and that each DNA segment from each individual in a pool is
equally likely to be read by the sequencing machinery.
Suppose that a read from the {\it i}'th pool is drawn from a DNA
segment covering our desired locus.
It is then expected that this read will contain the $B$ allele with probability
$q_i \equiv \frac{1}{2}\hat{\textbf{m}}_i \cdot \textbf{x}$,
where $\hat{\textbf{m}}_i$ is the {\it i}'th row of $\hat{M}$
(the $\frac{1}{2}$ pre-factor is due to the fact that both alleles are sequenced for each individual.)
The vector of frequencies of the $B$ allele, for each of the $k$ pools is therefore:
\be
\textbf{q}=\frac{1}{2}\hat{M}\textbf{x}
\ee

Had we been able to obtain a full and error-free coverage of the DNA present in the pool,
our measurements would have provided us with the exact value of $\textbf{q}$.
In practice a specific position is covered by a limited number of reads,
which we denote by $r$,
and the number of reads from the rare alleles $\textbf{z}$ out of the total
number of reads $r$ is binomially distributed $z_{i}\sim Binomial(r,q_i)$.
Generally, one can only control the expected number of reads covering a specific locus,
as also $r$ is considered a random variable. The main cause for variation in $r$ is the
 different amplification biases for different regions, which are effected
by properties such as a region's GC-content. The distribution of $r$ over different loci
depends on the experimental conditions, and was shown to follow a Gamma distribution
in certain cases \cite{Prabhu:01}. We adopt this assumption, draw $r$ for each locus
from a Gamma distribution $r \sim \Gamma(\frac{R}{L}, 1)$, and apply it to all $k$ pools.

This binomial sampling process provides measurements which are close, yet not identical to the
expected frequency of the rare allele $r q_i$, and these fluctuations are regarded as {\it sampling noise}.
Therefore the \cs problem formulation is given by (compare to Eq.~(\ref{eq:cs_problemFinal})):
\ber
\label{eq:CSopt}
\textbf{x}^* = \underset{\textbf{x} \in \{0,1,2\}^N}{argmin} \|\textbf{x}\|_{1} \quad {\mbox s.t.} \: \:
||\frac{1}{2} \hat{M}\textbf{x}-\frac{1}{r}\textbf{z}||_2 < \eps
\eer

{\it Adding noise factors.}
The model described so far assumes that no noise or bias exist in our setting,
besides sampling noise which is related to the limited number of reads.
In a more realistic scenario, we do expect additional noise factors to be present
due to imperfection in experimental procedures.
We have modeled these factors by adding two more types of noise: sequencing read errors
and errors in DNA preparation.

Read error models the noise factors introduced throughout the process of sequencing
by next generation techniques such as Illumina,
and reflect the fact that reads obtained from the sequencing machine may not match
the DNA molecule sampled.
This can be due to errors in certain bases present in the read itself,
mis-alignment of a read to a wrong place in the genome,
errors introduced by the PCR amplification process
(which are known to introduce base substitutions in the
replicated DNA~\cite{freeman1999quantitative}), or any other unknown factors.
All of these can be modeled using a single parameter $e_r$, which represents the
probability that the base read is different from the base of the measured sample's DNA at a given locus.
The resulting base can be any of the other three different nucleotides,
however we conservatively assume that the errors will {\it always} produce the
alternative allele $B$ (if, for example the reference allele
is `G' and the alternative allele is `T', we assume that all erroneous reads produce `T'.
In practice, some reads will produce `A' and `C', and these can be immediately discarded thus
reducing the effective error rate.)
The probability of observing $B$ at a certain read is therefore obtained by a convolution of the
frequency of $B$ alleles and the read error:
\be
\label{eq:q_vec}
\textbf{q} = (1-e_r) \frac{1}{2}\hat{M}\textbf{x} +
e_r (1-\frac{1}{2}\hat{M}\textbf{x})
\ee

The value of $e_r$ may vary as a function of the sequencing technology,
library preparation procedures, quality controls and alignment algorithms used.
Typical values of $e_r$, which represent realistic values for Illumina sequencing~\cite{Kao:01},
are in the range $e_r \sim\! 0.5\%\!- 1\%$.
We assume that $e_r$ is known to the researcher,
and that it is similar across different lanes. In this case, one can correct for the convolution
in Eq.~(\ref{eq:q_vec}) and obtain the following problem:
\ber
\label{eq:CSopt_noise}
\textbf{x}^* = \underset{\textbf{x} \in \{0,1,2\}^N}{argmin} \|\textbf{x}\|_{1} \quad {\mbox s.t.} \: \:
||\frac{1}{2}\hat{M}\textbf{x}-( \frac{1}{r}\textbf{z} - e_r ) / (1-2 e_r)||_2 < \eps
\eer
Hence the measurement vector in our problem equals $\textbf{y} = ( \frac{1}{r}\textbf{z} - e_r ) / (1-2 e_r)$ and the
sensing matrix is $\frac{1}{2} \hat{M}$.
If $e_r$ is unknown, one may still estimate it, for example by running one lane with a single
region on a single individual with known genotypes. Alternatively, we show in Appendix \ref{app:unknown_p_err}
how to incorporate the estimation of the read error term $e_r$ within our \cs framework from the
overlapping pooled sequence data. The noise factors described thus far (including sampling noise and
read errors) resembles the one proposed previously in \cite{Prabhu:01}.

Finally we add to our model one more source of noise, namely DNA preparation errors (\ppt errors.)
This error term reflects the fact that in an experimental setting it is hard to
obtain exactly equal amounts of DNA from each individual.
The differences in the actual amounts taken result in noise in the measurement matrix $M$.
While $M$ is our original zero-one Bernoulli matrix, the actual measurement matrix $M'$ is obtained
by adding \ppt errors to each non-zero entry.
Hence, the true mixture matrix is $M' \equiv M+D$,
where the \ppt matrix $D$ adds a centered Gaussian random variable to each non-zero entry of $M$:
\ber
d_{ij} \sim \left\{\begin{array}{ll}
N(0,\sigma^2) \hspace{1cm} \mbox{if} \:\: m_{ij}=1 \\
\equiv 0 \hspace{1.6cm} otherwise.
\end{array} \right.
\eer

We consider values of $\sigma$ in the range of $0 - 0.05$
reflecting up to $\sim\! 5\%$ average noise on the DNA quantities of each sample.
The matrix $M'$ is unknown and we only have access to $M$,
hence the form of Eq.~(\ref{eq:CSopt_noise}) in this case is unchanged.
$M'$ takes effect indirectly by modifying $\textbf{q}$,
which effects $\textbf{z}$ the actual number of reads from the rare allele.
As opposed to a classic \cs problem in which the sensing matrix is usually assumed
to be known exactly, \ppt effectively introduces noise into the matrix itself.
We study this effect of \ppt errors in Section~\ref{sec:Results},
and show that a standard \cs approach is robust to such noise.

{\it Targeted region length and coverage considerations.}
The expected number of reads from a certain locus is determined by the
total number of reads in a lane $R$ and the number of loci covered in a single lane $L$,
and is given by $E[r] = \frac{R}{L}$ (the actual number of reads from each locus $r$ follows a
Gamma distribution with mean $\frac{R}{L}$.)

$L$ is determined by the size of the regions and the number of SNPs of interest in a given study,
and by the ability of targeted selection techniques \cite{albert2007direct,gnirke2009solution,Ng:01}
to enrich for a given small set of regions.
We consider $L$ as a parameter and study its effect on the results.
When we treat different isolated SNPs, $L$ indeed represents the number of SNPs we cover, as each read covers one SNP.
When interested in contiguous genomic regions, however, $L$ should be interpreted
as the length of the target region in {\it reads}, rather than nucleotides, since
each read covers many consecutive nucleotides. Therefore one should multiply $L$ by the read length.
For example, if our reads are of length $50$ nucleotides, and $L$ is taken to be $100$,
we in fact cover a genomic region of length $5kb$.

$R$ is defined as the number of reads which were successfully
aligned to our regions of interest. It is mostly determined by the sequencing technology,
and is in the order of millions for modern sequencing machines.
$R$ is also greatly influenced by the targeted selection techniques used,
and since these are not perfect, a certain fraction of reads might not originate from the desired regions
and is thus `wasted'. The total number of reads varies according to experimental protocols, read length and
alignment algorithms but is typically on the order of a few millions.
Throughout this paper we have fixed $R$ to be $R = 4000000$, representing a rather conservative estimate
of a modern Illumina genome analyzer's run (e.g. compare to~\cite{mardis2008impact}),
and also assuming that targeted selection efficiency is very high (it is reported to be
$\sim90\!\%$ in \cite{gnirke2009solution}.)
Other values of $R$ may be easily dealt with using our simulation framework,
thus adapting to a particular researcher's needs.

Another important and related parameter is the {\it average coverage per individual per SNP},
denoted $c$, which is given by:
\be
\label{eq:coverage}
c = \frac{R/L}{N/2} \Big({\equiv \frac{E[r]}{N/2}}\Big)
\ee

Our model does not directly use $c$ and it is provided merely as a rough estimate for
the coverage in Section~\ref{sec:Results}, as it can be easily interpreted and compared to
coverage values quoted for single sample sequencing experiments.
When the total number of reads in a pool $r$ is given, the actual coverage obtained for each person
in a pool has a distribution which is approximately $Binomial(r, 1/N_{pooled})$, where
$N_{pooled} \sim \frac{N}{2}$ is the number of individuals in the pool.
Therefore the average coverage per individual in a given pool is indeed approximately $c$.

\subsection{Example}
\begin{figure}[thb!]
\begin{center}
\begin{tabular}{cc}
\psfig{file=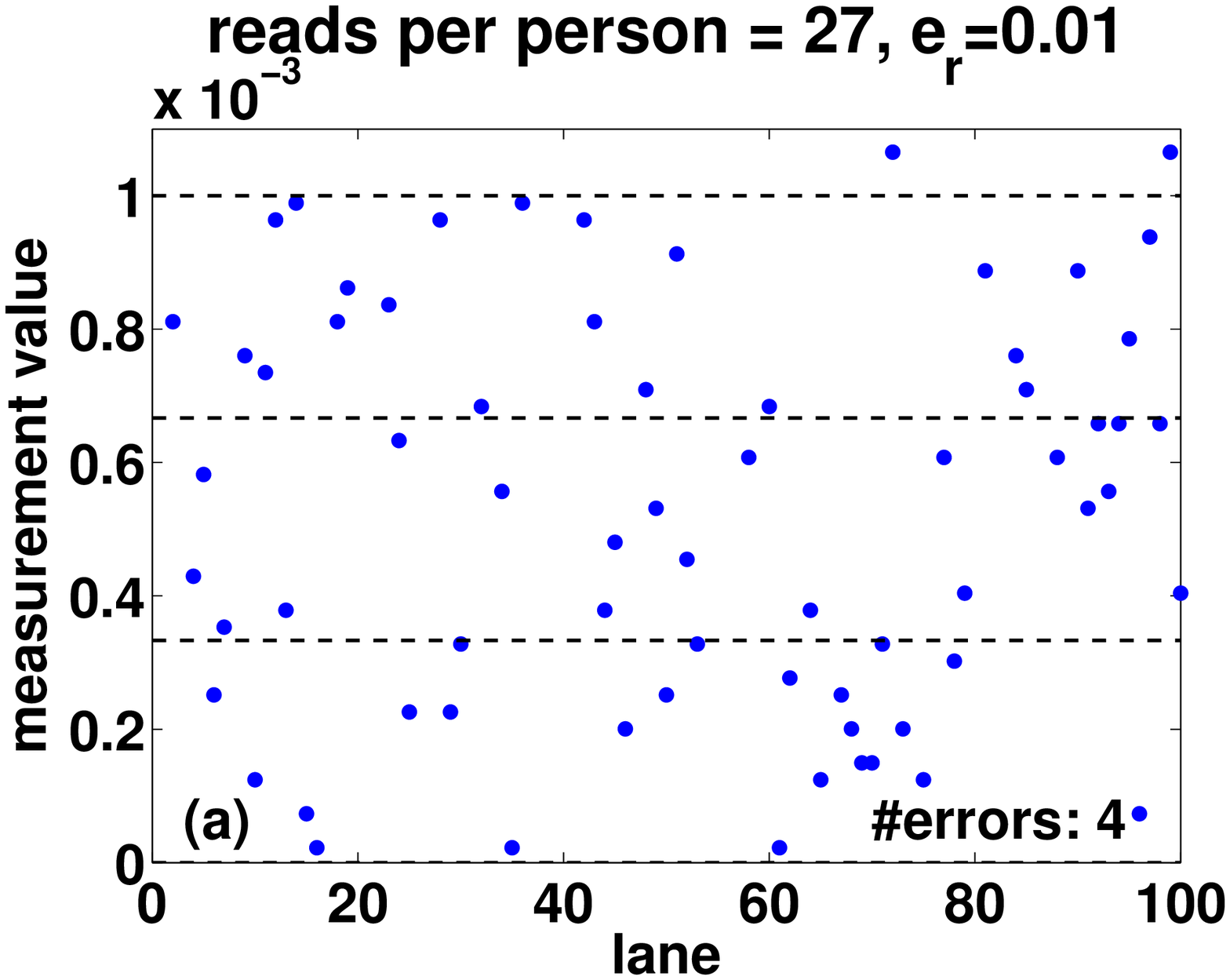,width=5.5cm} &
\psfig{file=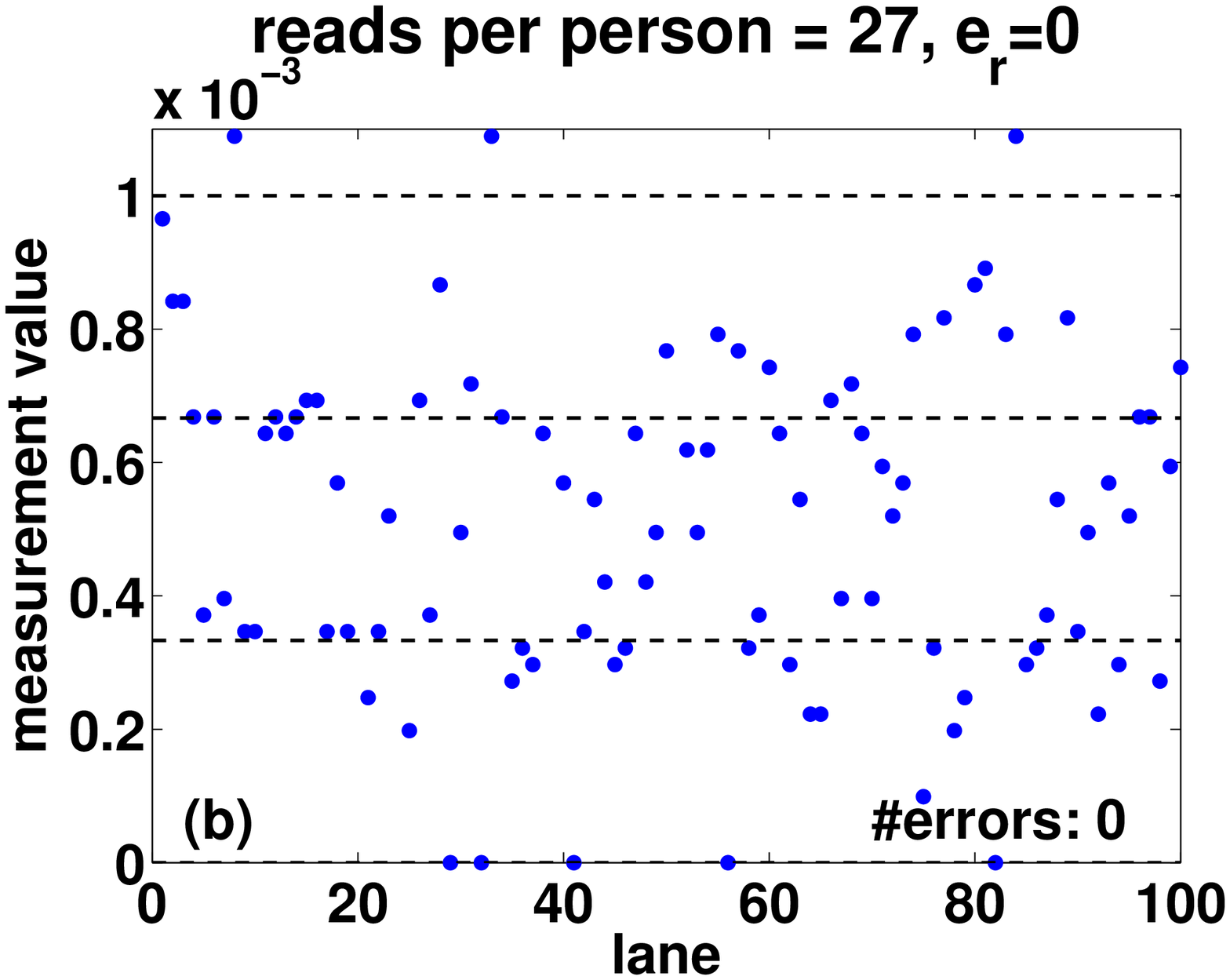,width=5.5cm}\\
\psfig{file=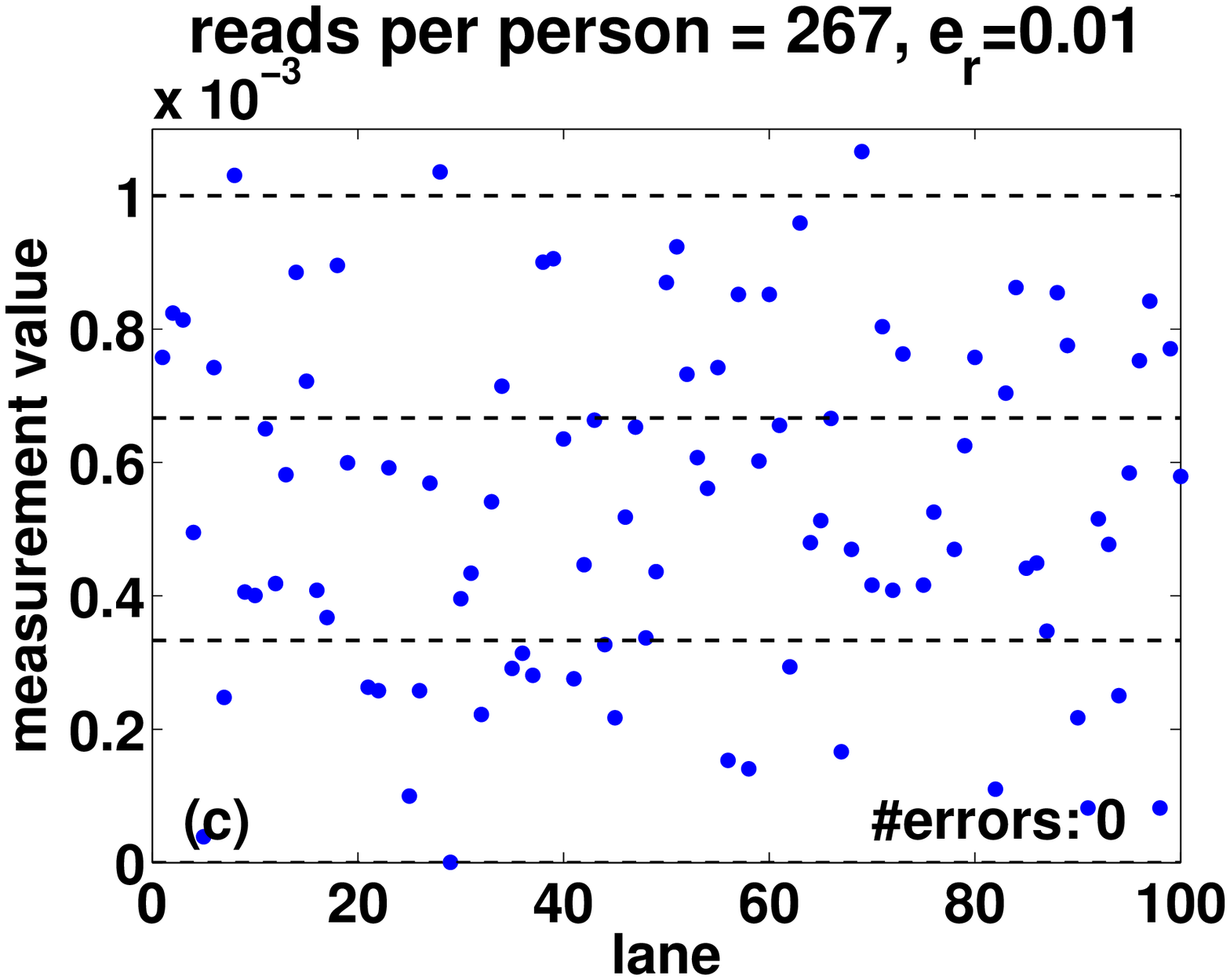,width=5.5cm} &
\psfig{file=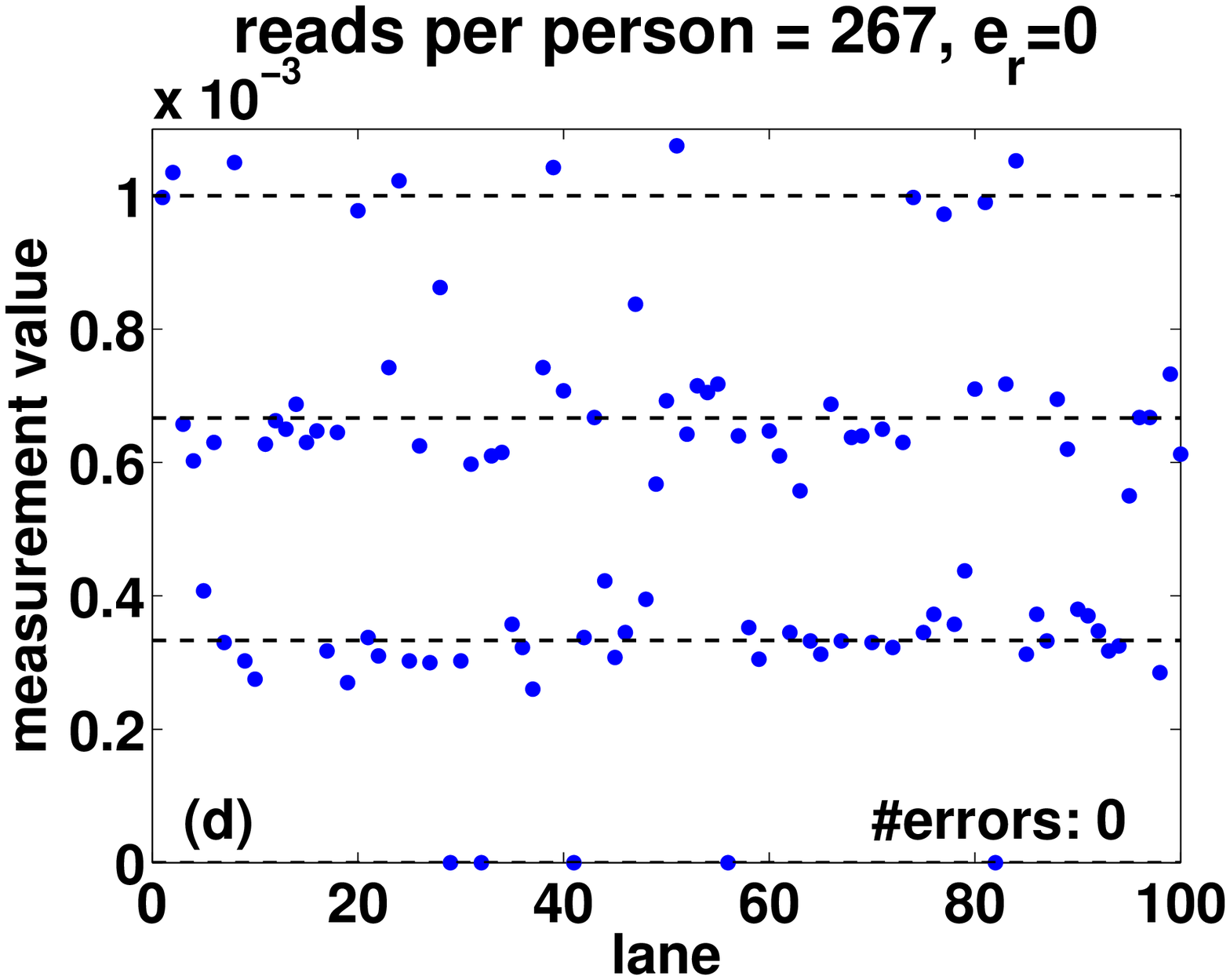,width=5.5cm}\\
\psfig{file=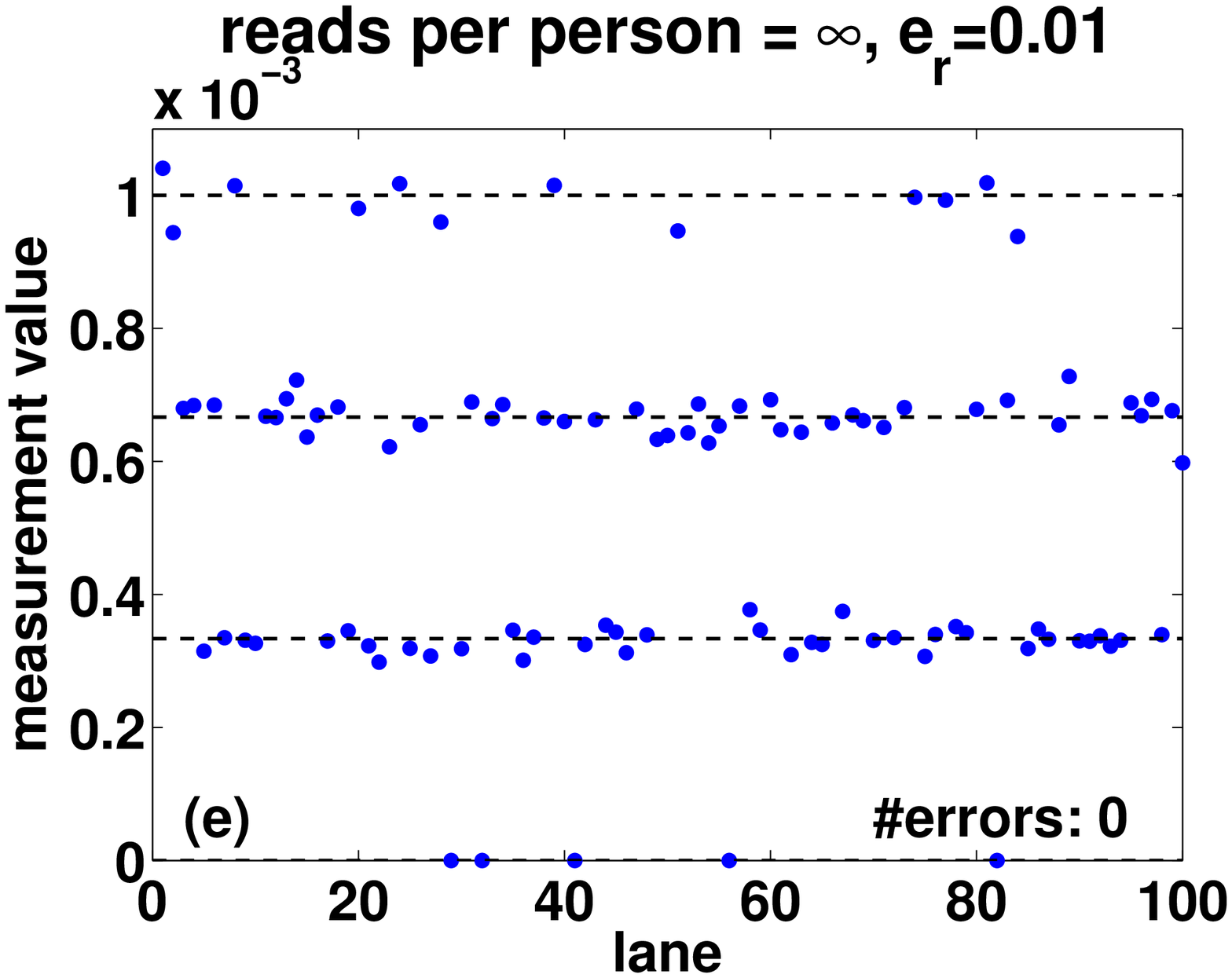,width=5.5cm} &
\psfig{file=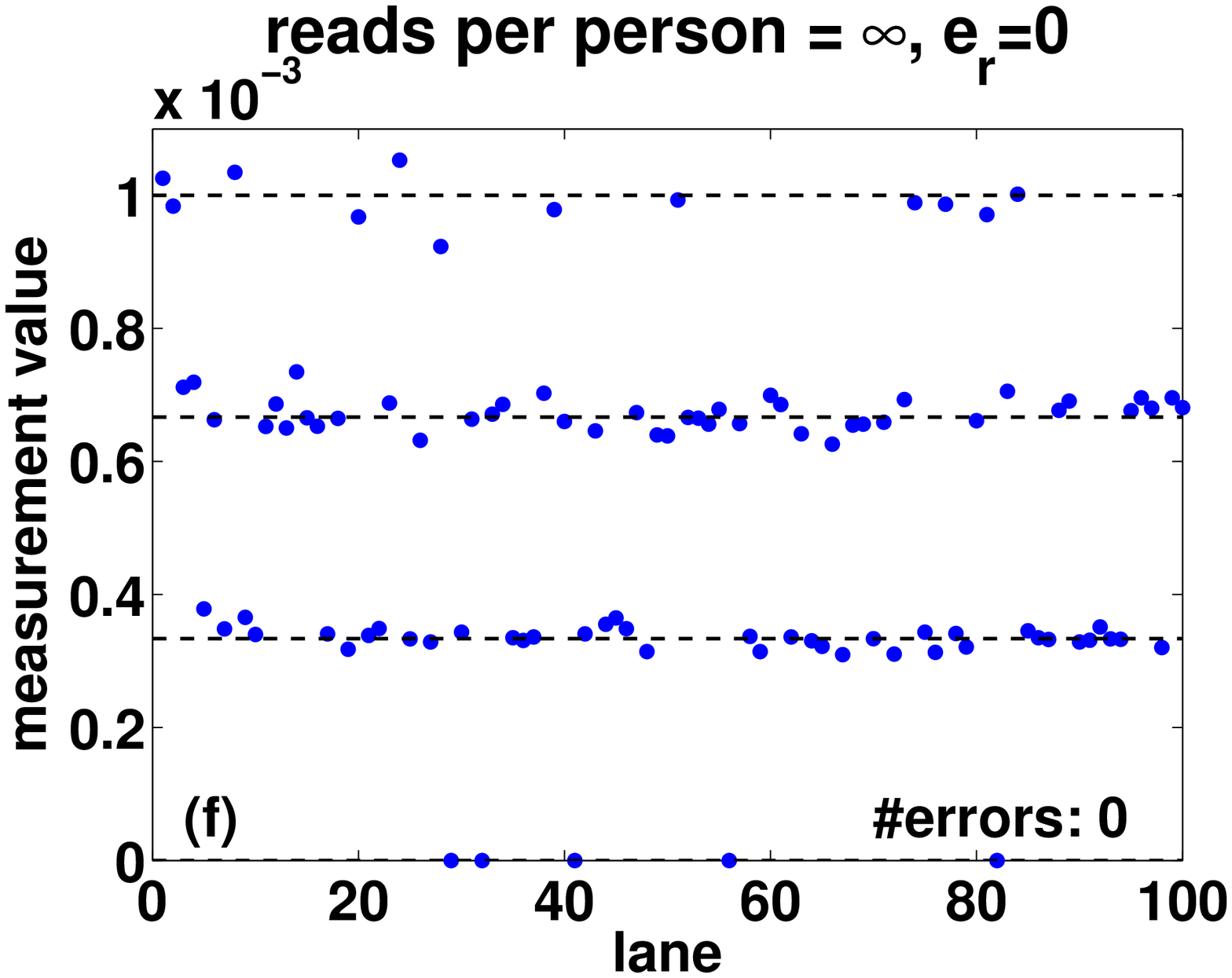,width=5.5cm} \\
\end{tabular}
\caption{The values measured for $100$ different pools,
for a specific case of $N=3000$ individuals and rare allele frequency $f=0.1\%$,
thus $3$ individuals carry the alternative allele.
Shown are the measured rare-allele frequencies in each lane, for different coverage levels,
($27$, $267$ and $\infty$ reads per pool), and different
values of the read error ($e_r = 0\%, 1\%$).
The data in all panels contain \ppt errors with $\sigma=0.05$.
The dashed lines represents the possible expected frequencies corresponding to
$0,1,2,3$ rare-allele carriers in a pool - these are the values
we would have obtained in the absence of read error, \ppt errors,
and assuming that each pool contains exactly $N/2$ individuals.
The coverage $r$ (i.e. number of reads) is the most dominant factor
causing deviations of the observed values from their expectancy.}
\label{fig:actualValues}
\end{center}
\end{figure}

In order to visualize the effect of the three noise factors, i.e., sampling noise, read errors and \ppt errors,
Fig.~\ref{fig:actualValues} presents the measured values $\textbf{y}$ in a specific scenario.
We simulate an instance of $N=3000$ individuals and rare allele frequency $f=0.1\%$,
tested over $k=100$ lanes.
Hence we have three heterozygotic carriers to be identified, and, in the absence of noise,
the measurement in each lane should display four levels,
which correspond to whether $0$, $1$, $2$ or $3$ of the carriers
are actually present in the specific pool.

In order to display the effect of sampling noise,
we consider three values for the average coverage $c$ , i.e.,
number of reads per individual per SNP:
$27$, $267$ and an infinite number of reads, which corresponds to zero sampling noise.
Each of these three values appears on a separate row in Fig.~\ref{fig:actualValues}.
The panels on the left hand side of Fig.~\ref{fig:actualValues} correspond to read error $e_{r}=1\%$,
while on the right hand side there is no read error at all.
The data in all panels contain \ppt errors with $\sigma=0.05$.
Each panel also displays the actual number of reconstruction errors in each case,
namely the Hamming distance between the correct vector $\textbf{x}$ and reconstructed vector
$\textbf{x}^*$ obtained by solving Eq.~(\ref{eq:CSopt_noise}).

The effect of sampling noise is clearly visible in Fig.~\ref{fig:actualValues}. An infinite
amount of reads (lower row; (e),(f)), causes the measurements to be very close to their expected frequency,
where slight deviations are only due to \ppt errors and the fact that the pool size is not exactly $N/2$.
For a moderate number of reads ($c=267$ - middle row; (c),(d)) the measurements follow the expected
frequency levels when there are no read errors (right panel (d)),
but this rough quantization completely vanishes for $e_{r}=1\%$ (left panel (c).)
However, reconstruction was accurate even in this case,
because our \cs formulation (Eq.~(\ref{eq:CSopt_noise})) takes these errors into account
and aggregate the information from {\it all} lanes to enable reconstruction.
When the number of reads per person is small ($c=27$ - upper row; (a),(b)),
the four levels disappear irrespective of the read error.
Reconstruction is still accurate in the absence of read errors (right panel (b)),
and there are $4$ errors in the reconstructed genotype vector ${\textbf{x}}^*$ when $e_{r}=1\%$ (left panel (a)),
which probably implies that sampling noise is too high in this case.
While a coverage of $27$ reads per person is overwhelmingly sufficient when sequencing
a {\it single} individual, it leads to errors in the reconstruction when pooling many individuals together.

\subsection{Reconstruction}
We use the Gradient Projection for Sparse Reconstruction algorithm (GPSR)~\cite{GPSR:01},
to solve the optimization problem~(\ref{eq:CSopt_noise}).
GPSR is designed to solve a slightly different, but equivalent,
formulation as in Eq.~(\ref{eq:CSopt_noise}):
\be
{\textbf{x}}^* = \underset{\textbf{x}}{argmin} \quad
||\frac{1}{2}\hat{M}\textbf{x}-( \frac{1}{r}\textbf{z} - e_r ) / (1-2 e_r)||_2 + \tau ||\textbf{x}||_1
\label{eq:cs_problem_GPSR}
\ee

where the parameter $\tau$ provides the trade-off between the equations
fit and the sparsity promoting factor, and is equivalent to specifying the maximal allowed error $\eps$.
It is often desirable in applications to let the parameter $\tau$ scale with $||M^{\textbf{T}}{\textbf{y}}||_{\infty}$
\cite{kim2007interior}, where in our case $\textbf{y}=( \frac{1}{r}\textbf{z} - e_r ) / (1-2 e_r)$
corresponds to the measurements. We have chosen to adopt this scaling throughout this paper, and have set
$\tau=0.01||\hat{M}^{\textbf{T}}{\textbf{y}}||_{\infty}$,
although experimentation with different values of $\tau$ did not alter results significantly.

GPSR outputs a sparse vector ${\textbf{x}}^*$ with a few non-zero real entries,
but does not use the fact that our variables are integers from the set $\{0,1,2\}$.
We therefore perform a post-processing step in order to obtain such a solution.
Simple rounding of the continuous results in ${\textbf{x}}^*$ may obtain such a vector\footnote{by `rounding'
here we mean reducing to the set $\{0,1,2\}$. Thus any negative number is `rounded' to zero and any number
larger than $2$ is `rounded' to $2$.}.
We chose a different post-processing scheme, which yields better performance:
we rank all non-zero values obtained by GPSR, and round the largest $s$ non-zero values,
setting to zero all other $N-s$ values to get the vector ${\textbf{x}}^{*s}$.
We then compute an error term $err_s \equiv ||\frac{1}{2}\hat{M} {\textbf{x}}^{*s} - \textbf{y}||_2$.
Repeating this for different values of $s$ we select the vector ${\textbf{x}}^{*s}$
which minimizes the error term $err_s$.
Thus the final solution's sparsity $s$ is always smaller or equal to the sparsity of the vector
obtained by GPSR.

\subsection{Utilizing Barcodes}
\label{sec:MethodsBarcodes}
In this section we describe how a \cs\!-based method can be combined with a barcoding strategy
resulting in improved performance. A barcode is obtained by attaching to the DNA in each sample a unique DNA
sequence of about $5$ additional nucleotides, which enables the unique identification of this
sample \cite{hamady2008error}.
Hence samples with different barcodes can be mixed together into a single lane, and
reads obtained from them can be uniquely attributed to the different samples.
In a pooled-barcodes design \cite{Erlich:01}, the DNA in each pool (as opposed to the DNA of
a specific individual) is tagged using a unique barcode. If $n_{bar}$ different
barcodes are available, we may apply $n_{bar}$ pools to a single lane and still identify
the pool from which each read originated (although not the specific individual.)

We utilize barcodes by increasing the number of effective lanes while the number of
reads per lane is decreased. The usage of $k$ lanes and $n_{bar}$ barcodes
is simply translated into solving problem (\ref{eq:CSopt_noise}) with $k \times n_{bar}$ pools
and $R/n_{bar}$ total reads per lane.
Barcodes can therefore be combined easily with our \cs framework so
as to improve efficiency. We did not try to estimate the relative cost of
barcodes and lanes as it may vary according to lab, timing and technology conditions.
We therefore solve the \cs problem for different $(k,n_{bar})$ combinations,
thus presenting the possible trade-offs.

\subsection{Simulations}
We have run extensive simulations in order to evaluate the performance of our approach.
Various parameter ranges were simulated,
where each set of parameters was tested in $500$ instances.
In each simulation we have generated an input genotype vector $\textbf{x}$,
applied measurements according to our mathematical model,
and have tried to reconstruct $\textbf{x}$ from these measurements.
In order to evaluate the performance of our approach,
one needs a measure of reconstruction accuracy, reflecting the agreement
between the input vector $\textbf{x}$ and the reconstructed \cs vector $\textbf{x}^*$
(even when executing the naive and costly approach of
sequencing each individual in a separate lane, one still expects possible disagreements between
the original and reconstructed vectors due to insufficient coverage and technological errors.)
Each entry $i$ for which $x_i$ is different from $x^*_i$ is termed a reconstruction error,
and implies that the genotype for a certain individual was not reconstructed correctly, yielding
either a false positive ($x_i = 0 \neq x^*_i$) or a false negative ($x_i \neq 0 = x^*_i$).
For simplicity, we have chosen to show a simple and quite restrictive measure of error:
we distinguish between two `types' of reconstructions - completely accurate reconstructions
which have {\it zero} errors, and reconstructions for which at least one error occurred.

A certain value of the problem's parameters (such as number of individuals, number of lanes, read error etc.)
is termed ``successful'' if at least $95\%$ of its instances (i.e. $475$ out of $500$)
had {\it zero} reconstruction errors,
namely {\it all} individual genotypes were reconstructed correctly.
Thus, even when testing for a few thousand individuals,
we require that none of the reference allele carriers will be declared as a rare allele carrier.
In particular, this requirement guarantees that the False-Discovery-Rate of
discovering rare-allele carriers will not exceed $0.05$.

Performance is then measured in terms of $N_{max}$,
defined as the maximal number of individuals which allow for a ``successful'' reconstruction,
for certain values of the problem's parameters.

\section{Results}
\label{sec:Results}

To explore the advantages of applying \cs for efficiently identifying
carriers of rare alleles, we performed various computer simulations of the experimental
procedure described in Section~\ref{sec:methods}.

In each instance of the simulations we set the following parameters:
The number of individuals grouped together $N$ varied between $100$ and $20000$.
Rare allele frequency $f$ was chosen to be $0.1\%$, $1\%$ and $2\%$,
thus in each instance we randomly select $s = Nf$ carriers, and this determines our
input vector $\textbf{x}$. Since rare allele frequency is low,
we mostly consider the case of a heterozygous allele ($AB$),
hence $\textbf{x}$ is a binary vector, with $1$'s marking the carriers\footnote{Assuming a given locus follows
HW equilibrium, the expected frequency of homozygous rare allele carriers is $(1-\sqrt{1-f})^2 \sim f^2/4$
which is extremely low.}. The case of a homozygous alternative allele ($BB$) is presented
in Section~\ref{sec:lanesAndNmax1Per},
thus in this case $\textbf{x}$ can also contain the value $2$.
The number of lanes $k$ varied between $10$ and $500$,
and $L$, the number of targeted loci on the same lane was $1$, $10$, $100$ and $500$ (corresponding
to  targeted regions of length $100$ base-pairs to $50kb$, assuming each read is of length $100$),
which leads to different levels of coverage and sampling noise.
The other noise factors are kept fixed, with read error $e_{r}=1\%$
and \ppt errors $\sigma=0.05$, unless specified otherwise.

In Section~\ref{sec:lanesAndNmax} we estimate the performance of \cs given all relevant
noise factors, while in Section~\ref{sec:noiseEffect} we evaluate the individual effect of
each of the three noise factors, i.e., sampling noise, read errors and
\ppt errors. Section~\ref{sec:Msqrt} shortly presents the effect of using
a different sensing matrix in which only $\sim\!\sqrt{N}$ individuals participate
in each pool, instead of $\sim\!N/2$.
Finally, in Section~\ref{sec:resultsBarcodes} the effect of combining barcodes
and \cs is presented.

\subsection{Performance of the `standard' experimental setup}
\label{sec:lanesAndNmax}
Figures \ref{fig:lanesAndNmax1Pro},\ref{fig:lanesAndNmax1Per2Per} present $N_{max}$ as a
function of $k$, for different numbers of SNPs sequenced together on the same lane.
The case $f=0.1\%$ displays a different behavior than $f=1\%$ and $f=2\%$
and is considered separately.

\subsubsection{$f=0.1\%$}
\begin{figure}[thb!]
\begin{center}
\begin{tabular}{cc}
\epsfig{file=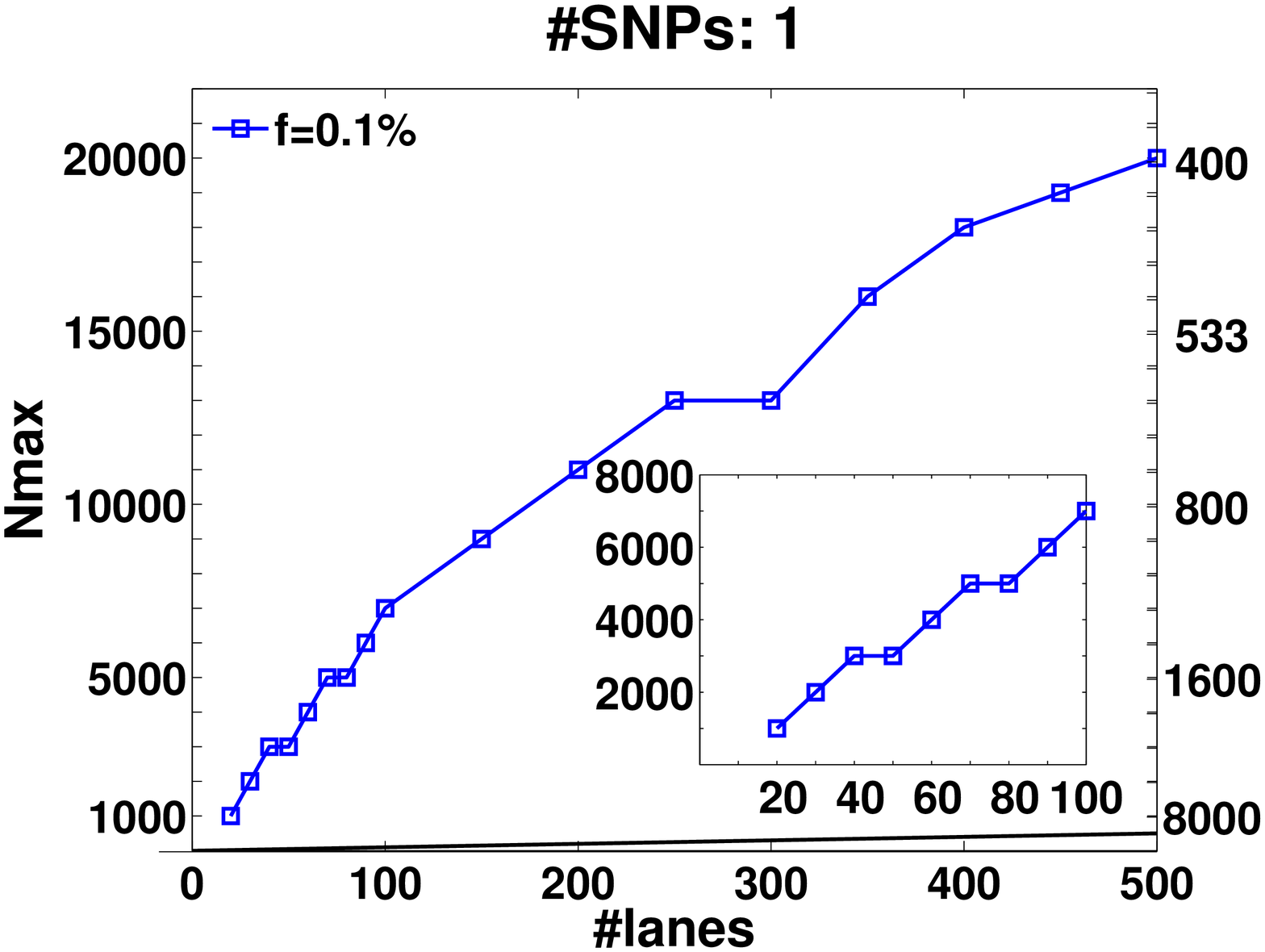,width=6cm} &
\epsfig{file=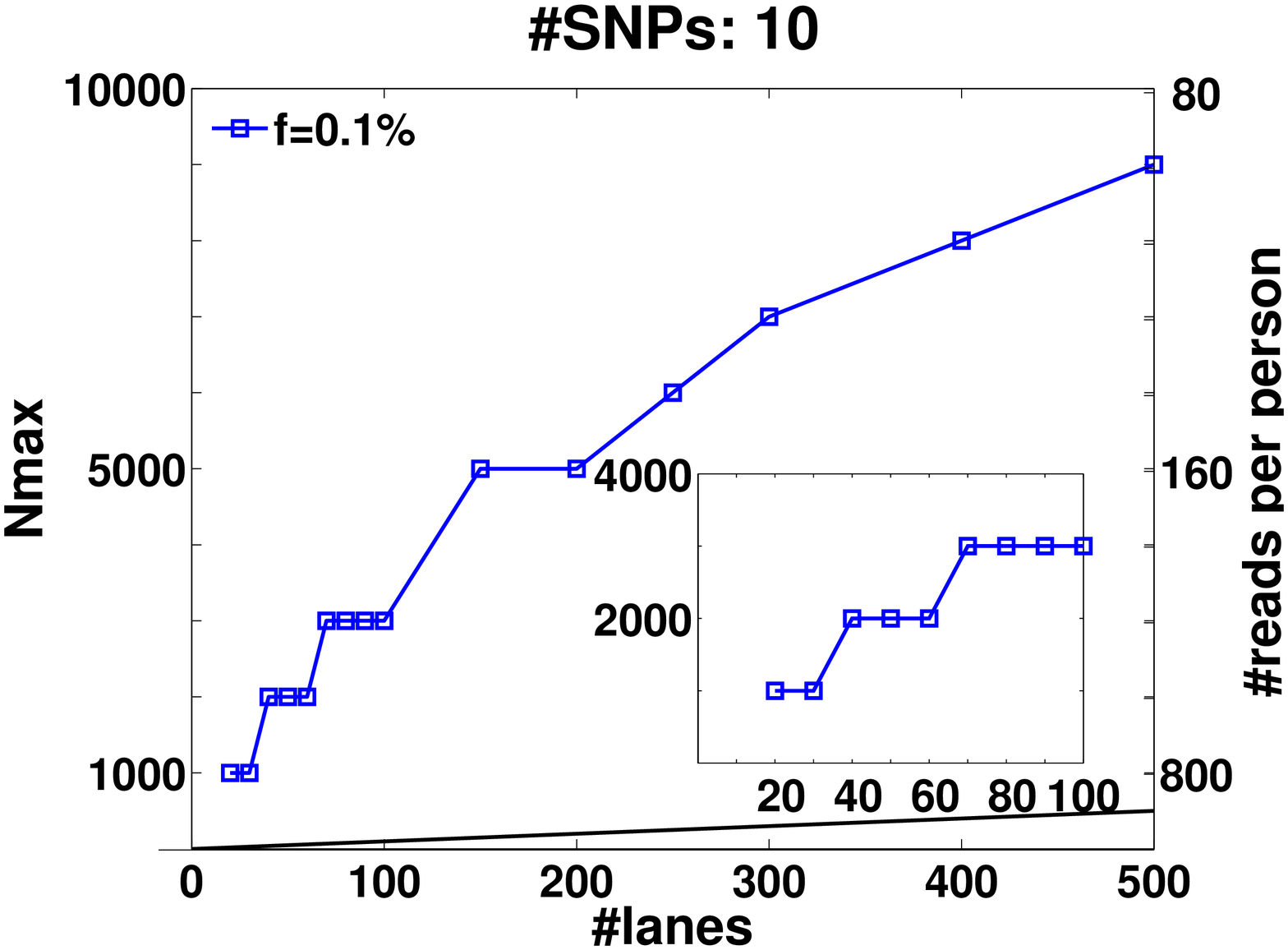,width=6cm} \\
\epsfig{file=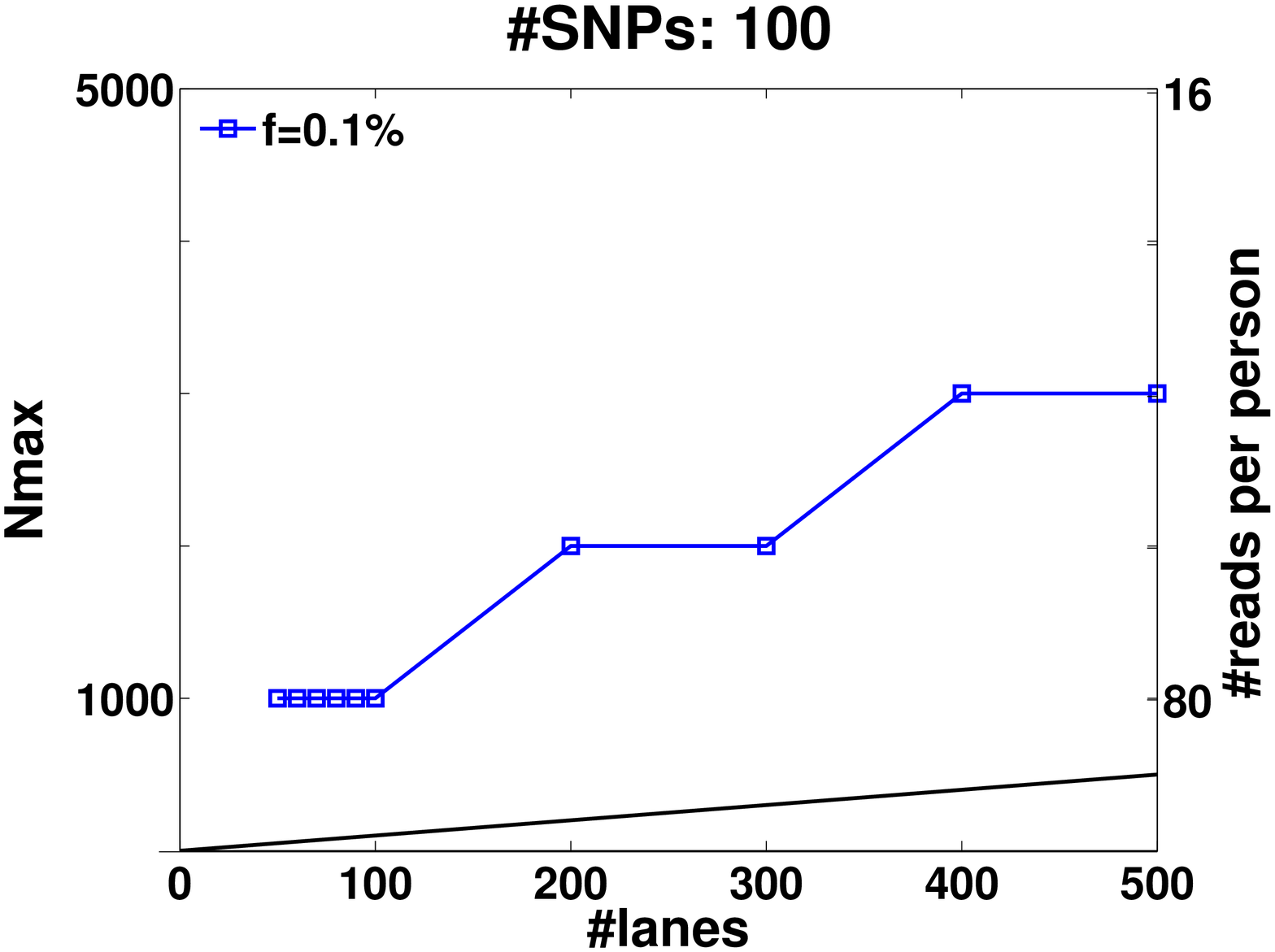,width=6cm} &
\epsfig{file=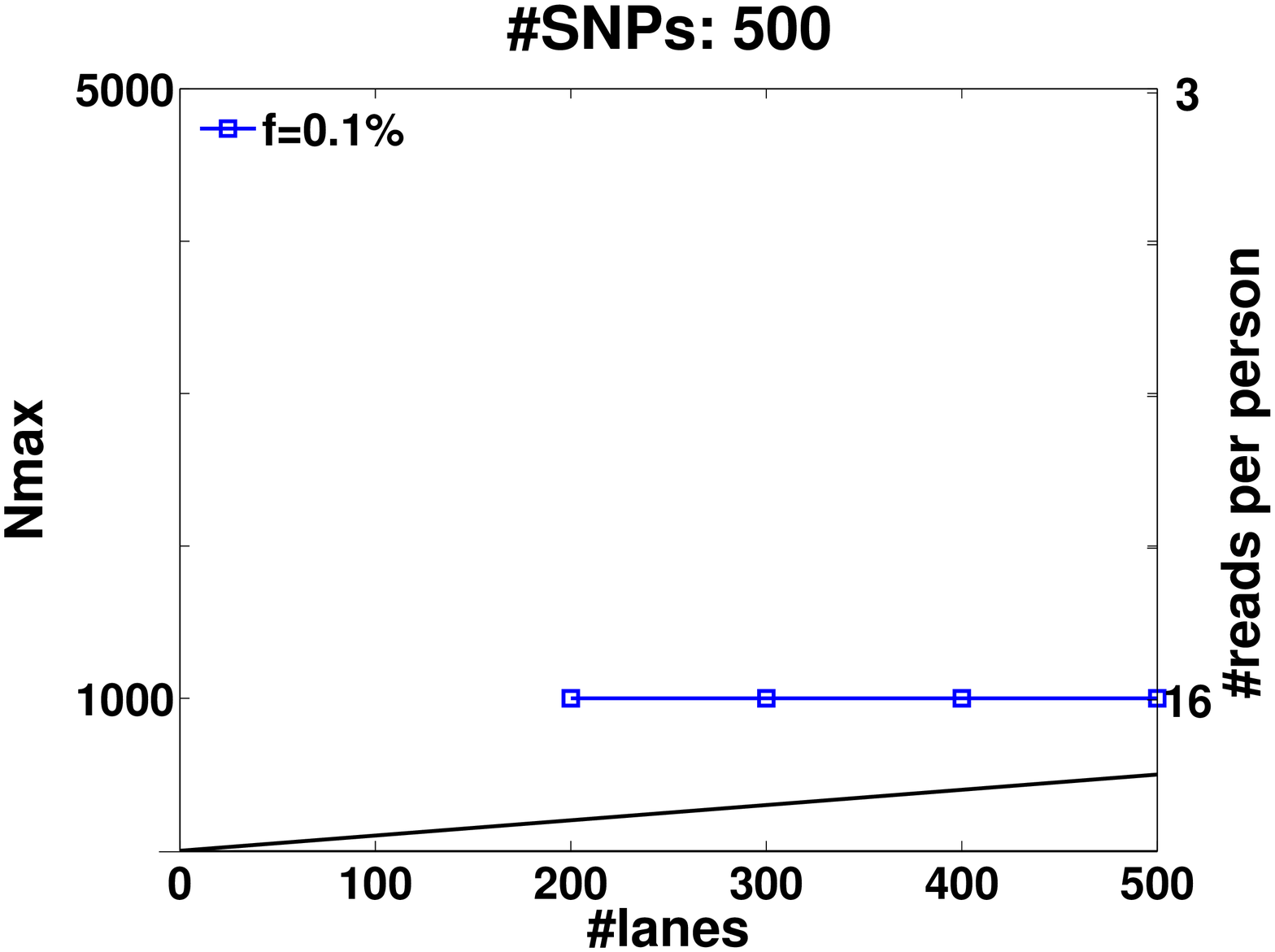,width=6cm}
\end{tabular}
\caption{The maximal number of individuals $N_{max}$
which allow for a ``successful'' reconstruction
as a function of the number of lanes used,
for different numbers of loci treated simultaneously.
A ``successful'' reconstruction means that for a certain set of parameters
at least 475 out of 500 simulations yield {\it zero} reconstruction errors.
The black line is simply the line $y=x$,
demonstrating the performance of the naive approach of using a
single lane per sample.
The vertical right axis displays the corresponding average coverage $c$ for every
value of $N_{max}$, obtained via Eq.~(\ref{eq:coverage}).
The bottom-right panels in (a) and (b) are zooming in on the region where the number
of lanes is small, which is at present the most realistic scenario
(in panel (c) $N_{max}$ was constant for low numbers of lanes.)
The values of $N_{max}$ in this case were taken in units of $1000$ individuals,
which correspond to single carriers. Cases which appear to be missing,
e.g. $k<200$ for $L=500$ simply mean that $N_{max}<1000$.}
\label{fig:lanesAndNmax1Pro}
\end{center}
\end{figure}

The advantages of \cs appear most dramatically in the case of rare alleles,
e.g., for $f=0.1\%$ in Fig.~\ref{fig:lanesAndNmax1Pro}.
Each panel in Fig.~\ref{fig:lanesAndNmax1Pro} presents $N_{max}$ as a function of $k$,
for different numbers of SNPs $L$.
The number of rare-allele carriers tested in this case were $1,2,\ldots,20$,
leading to $N=1000,2000,\ldots,20000$. The vertical right axis displays
the corresponding average coverage $c$, obtained via Eq.~(\ref{eq:coverage}).
The thick black line in each figure is simply the line $y=x$,
demonstrating the performance of the naive approach of using a single lane per sample.

When the number of available lanes is large, we can successfully identify the carriers in groups of
up to $9000$ or $20000$ individuals, for $k=500$ lanes,
and $L=10$ or $L=1$, respectively (Fig.~\ref{fig:lanesAndNmax1Pro}(a,b).)
In case the number of available lanes is small we can still identify a
single carrier out of $1000$ individuals with merely $k=20$ lanes, for $L=1$ and $L=10$.
(inset in Fig.~\ref{fig:lanesAndNmax1Pro}(a,b).)
With $k=30,40$ lanes, we can identify $2$ or $3$ carriers in a group of
$2000$ or $3000$, for $L=1,10$, respectively.

As is evident from the four panels of Fig.~\ref{fig:lanesAndNmax1Pro}, $N_{max}$
decreases as a function of $L$. For example, $500$ lanes are sufficient to deal
with $20000$ individuals for $L=1$, but only with $1000$ individuals for $L=500$.
This results from insufficient coverage which causes an increase in sampling noise.
Increasing the number of lanes can overcome this under-sampling as the value of $N_{max}$ increases
almost linearly with $k$ in most cases.

\begin{figure}[thb!]
\begin{center}
\psfig{file=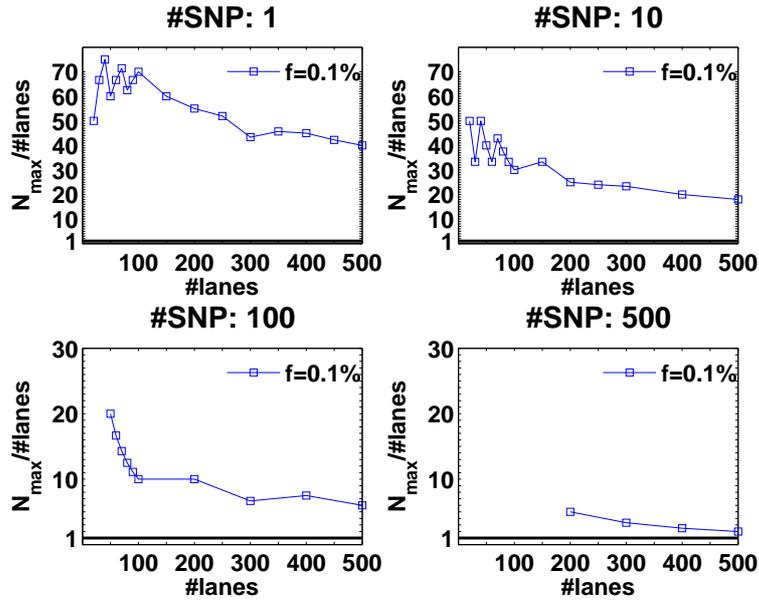,width=10cm}
\caption{Efficiency score for our approach. The ratio $N_{max}/k$ in our
approach and the number treated in the naive approach (equal to the number of lanes.)
This represents the ratio of saved resources (lanes.)
Efficiency is highest when a few lanes are used, and decreases gracefully as we use more
and more lanes.
Efficiency is highest when the targeted number of loci is small,
as in this case each lane provides very high coverage.}
\label{fig:lanesAndNmax1ProES}
\end{center}
\end{figure}

In order to quantify the advantage of applying \cs we define an ``efficiency score'',
presented in Fig.~\ref{fig:lanesAndNmax1ProES}, which is simply $N_{max}/k$,
i.e., the number of individuals for which reconstruction can be performed
using the \cs approach for a given number of lanes,
divided by the number of individuals which can be treated using the
naive one-individual-per-lane approach.
Therefore, the higher the score, the more beneficial it is to apply \cs\!\!.
The black line in each plot has a value of $1$ which corresponds to
the naive scenario of one individual per lane.
When considering up to $L=100$ SNPs, the efficiency score is around or above $10$,
and in some cases is as high as $70$.

\begin{figure}[thb!]
\begin{center}
\psfig{file=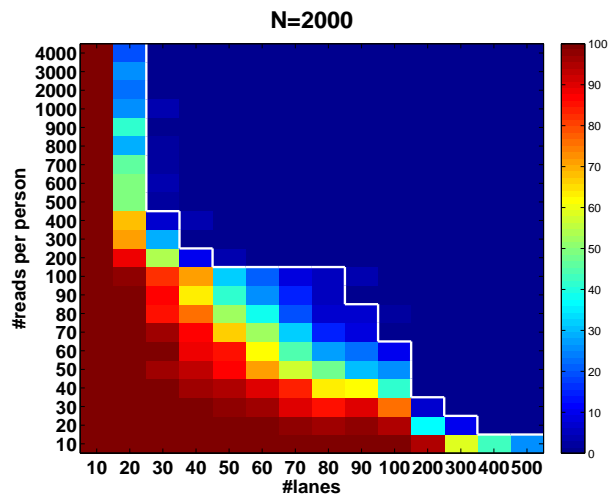,width=8cm}
\caption{Performance as a function of the number of lanes $k$ and the average coverage $c$.
Shown is the percentage of runs in which (even a single) reconstruction error occurred.
A rather sharp transition is shown, where above the white line we achieve a completely
accurate reconstruction in at least $95\%$ of the simulations.
Notice the non-linearity of scale in both axes.}
\label{fig:lanesAndNmax1ProLaneAndNmax}
\end{center}
\end{figure}

The axis on the right hand side of Fig.~\ref{fig:lanesAndNmax1Pro} displays the
average number of reads per person, i.e., average coverage $c$, for the relevant $N_{max}$.
One important question is related to the optimal number of reads
which allows for successful reconstruction: the smaller the coverage
the more SNPs we can test on the same lane, yet we are more prone to (mostly sampling) noise.
However, one can overcome the effects of low coverage by increasing the number of lanes, 
hence it is interesting to test the performance for each combination of
coverage $c$ and number of lanes $k$.
In Fig.~\ref{fig:lanesAndNmax1ProLaneAndNmax} we present this performance for $N=2000$ individuals and $f=0.1\%$.
For each pair of coverage and number of lanes $k$ we color code the percentage of instances for which there
were errors in \cs reconstruction.
An improvement in performance may be achieved both by increasing
the coverage and by increasing the number of lanes.
The white line marks the $95\%$ accuracy threshold.
The transition between ``successful'' and ``unsuccessful'' reconstruction is rather sharp.
For low coverage, e.g., lower than $100$ reads per person, a very high number of lanes is
needed in order to overcome sampling noise.

\newpage
\subsubsection{$f=1\%$, $f=2\%$}
\label{sec:lanesAndNmax1Per}
\begin{figure}[thb!]
\begin{center}
\begin{tabular}{cc}
\epsfig{file=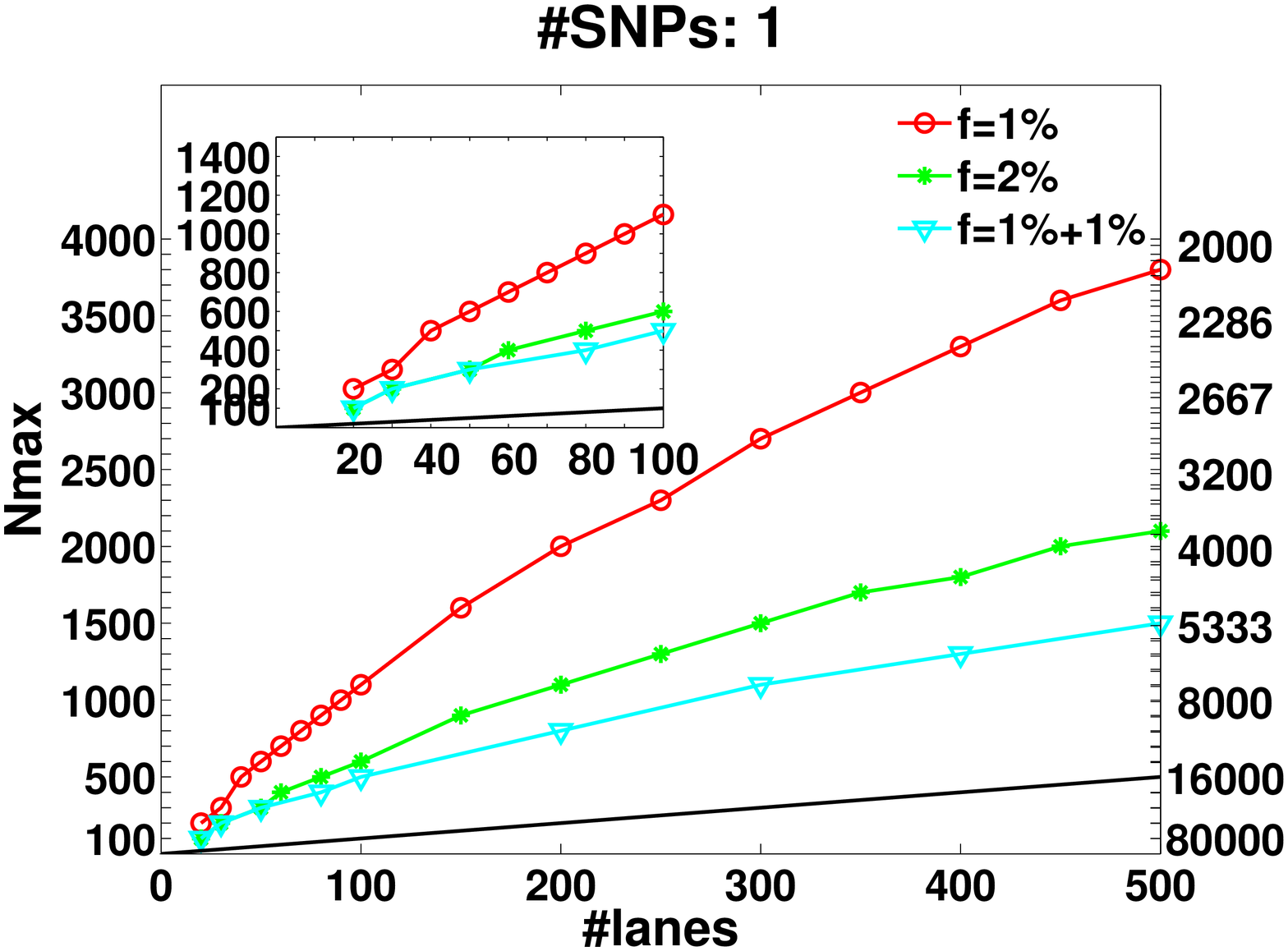,width=6cm} &
\epsfig{file=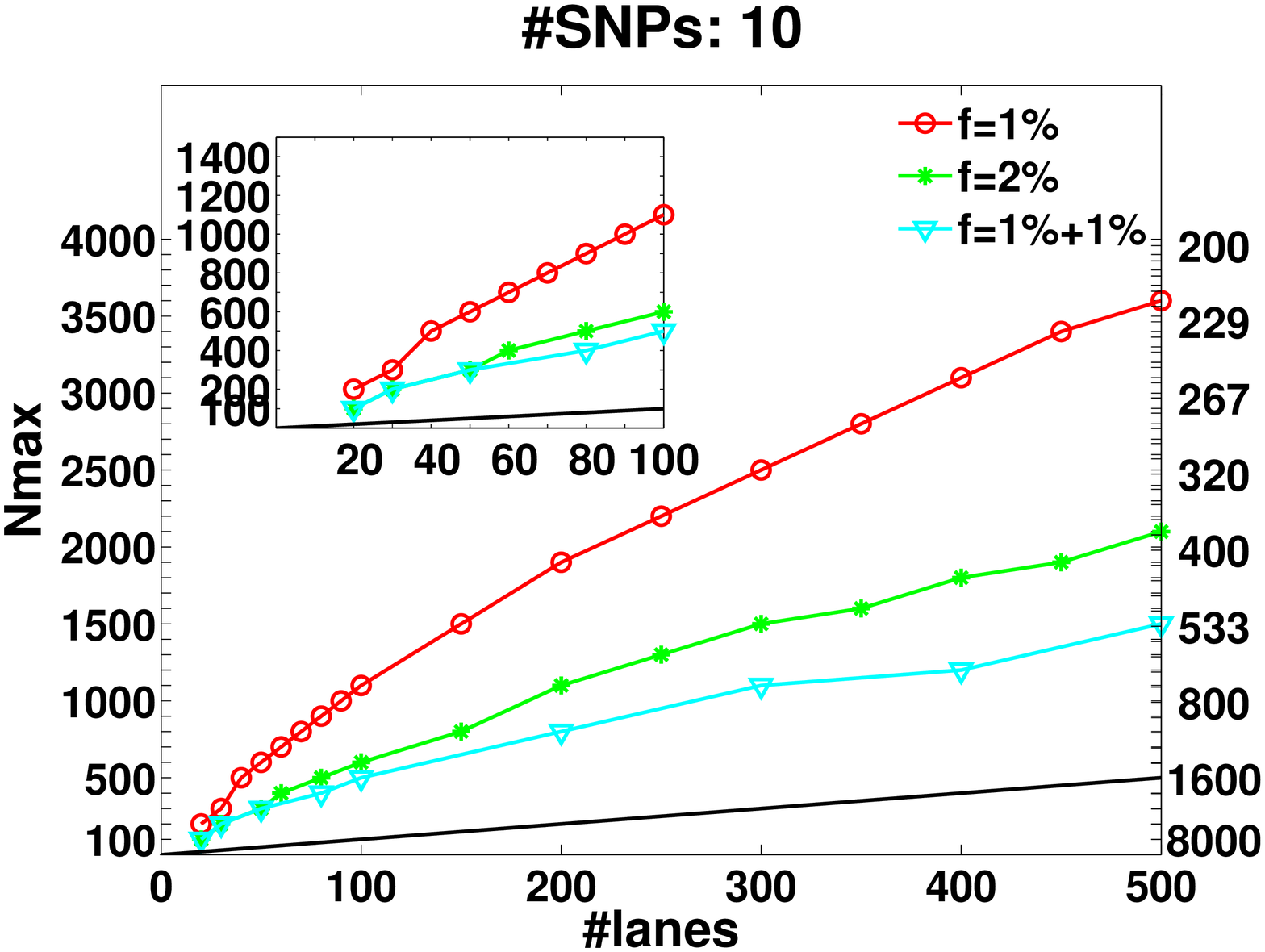,width=6cm} \\
\epsfig{file=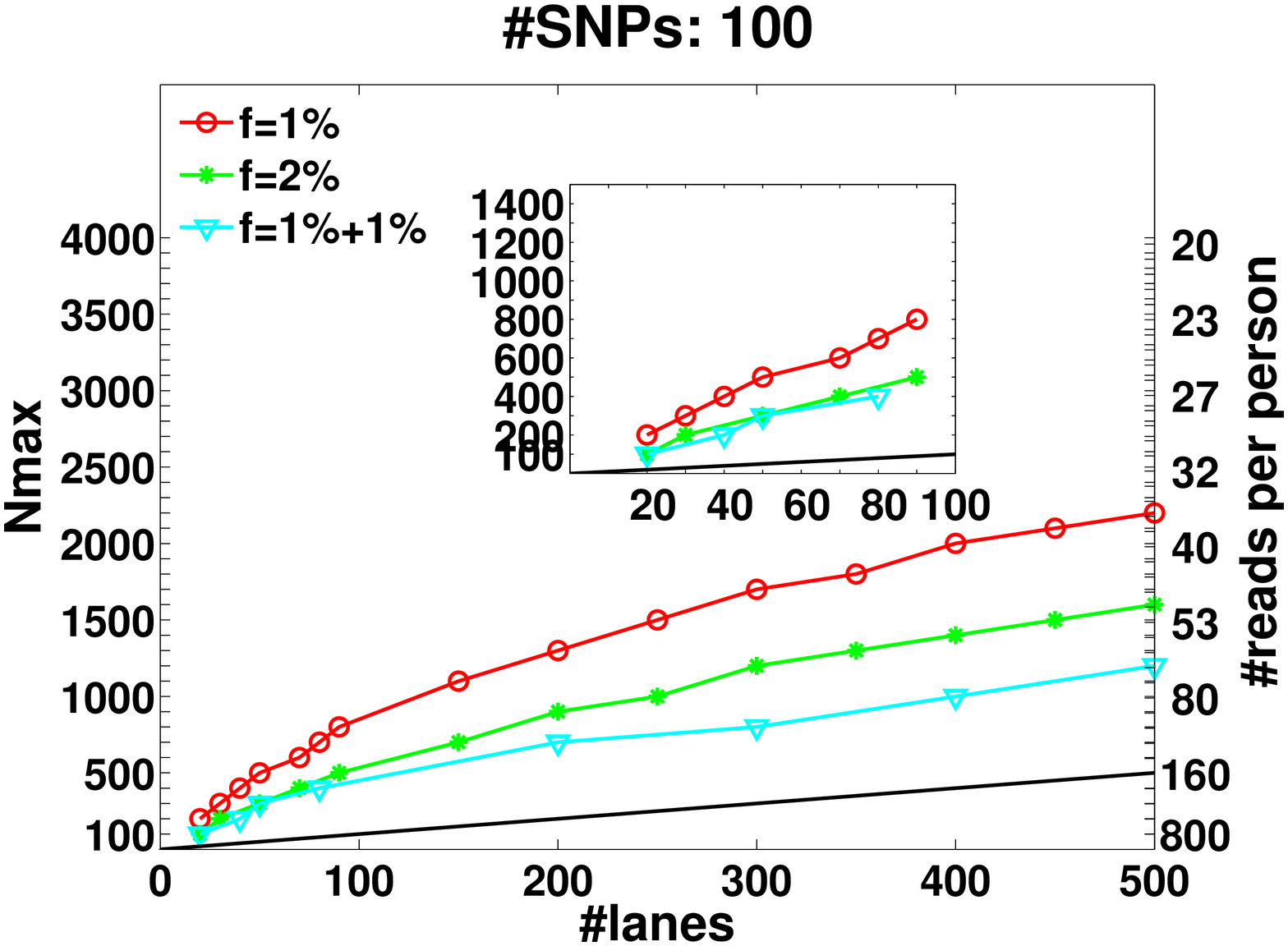,width=6cm} &
\epsfig{file=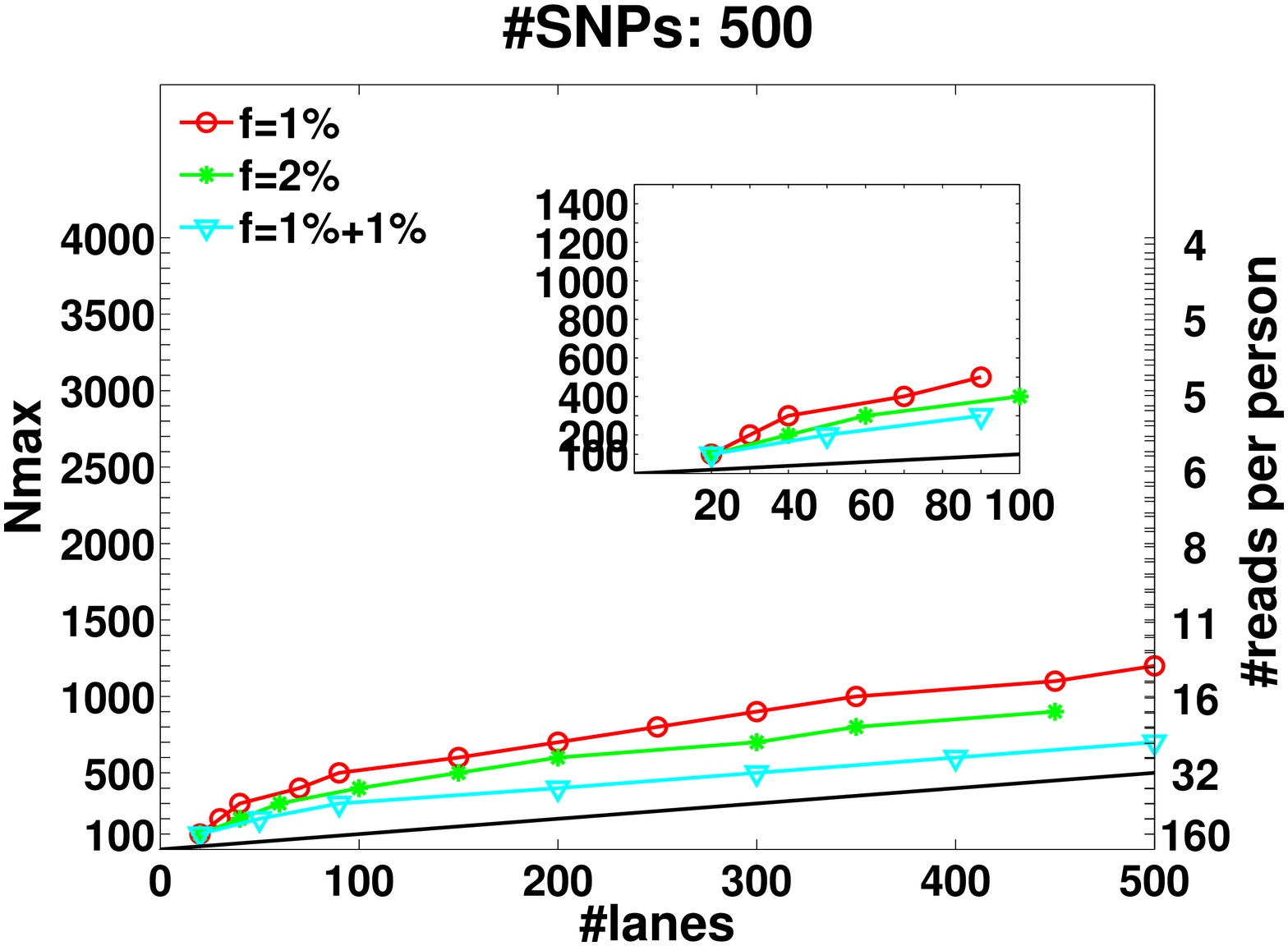,width=6cm}
\end{tabular}
\caption{The maximal number of individuals $N_{max}$ as a function of the number of lanes.
Similar to Fig.~\ref{fig:lanesAndNmax1Pro} but with a higher
frequency for the rare allele, $f=1\%$ and $f=2\%$.
The number of individuals $N_{max}$ achieved decreases as we increase the
rare allele frequency, but we are still able to treat a much larger sample size
than the naive approach. For example, one can use $40$ lanes
for $L=100$ and recover $4$ rare-allele carriers out of $400$ ($f=1\%$, zoomed-in view in panel(c).)
The case $f=1\%+1\%$ corresponds to $1\%$ $AB$ and $1\%$ of $BB$
alleles.}
\label{fig:lanesAndNmax1Per2Per}
\end{center}
\end{figure}

\begin{figure}[thb!]
\begin{center}
\psfig{file=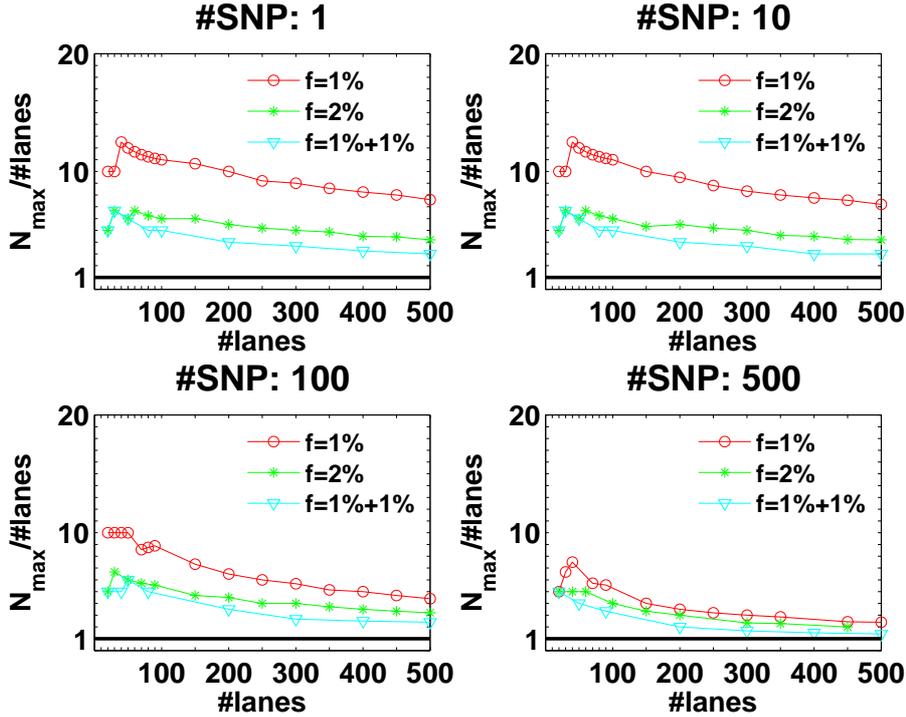,width=12cm}
\caption{Efficiency score of our approach. Similar to Fig.~\ref{fig:lanesAndNmax1ProES} but with
rare allele frequency of $1\%$ and $2\%$. As expected, efficiency is decreased as rare allele
frequency increases. We still reach up to $13X$ and $7X$ improvement over the naive approach for
rare allele frequencies of $1\%$ and $2\%$ respectively.}
\label{fig:lanesAndNmax1Per2PerES}
\end{center}
\end{figure}

Figure~\ref{fig:lanesAndNmax1Per2Per} presents the results for $f=1\%$ and $f=2\%$.
In this case the values of $N$ tested were $100,200,\ldots,4000$ (no successful reconstruction according
to our criteria was achieved for $N > 4000$.)
The resulting $N_{max}$ is lower than for the case of $f=0.1\%$, although still much higher than in
the naive approach. Results for $L=1$ are similar to those of $L=10$, namely increasing the coverage
does not improve performance significantly in this case.
The differences between results for $f=1\%$ and $f=2\%$ are rather small.
The ``efficiency score'' in this case is lower (see Fig.~\ref{fig:lanesAndNmax1Per2PerES}),
and is around $5$, still offering a considerable saving compared to the naive approach.

All former simulations considered the case of identifying carriers of a heterozygous allele ($AB$).
In order to study the possibility of also identifying homozygous alternative alleles ($BB$) via \cs
we simulated the following case: $1\%$ of the individuals are $BB$ in addition to $1\%$ which are $AB$
(this gives a vastly higher frequency of $BB$ than is expected to be encountered in practice,
and was taken as an extreme case to test the robustness of our reconstruction results.)
The results are marked as ``$1\%+1\%$'' in Figs.~\ref{fig:lanesAndNmax1Per2Per},\ref{fig:lanesAndNmax1Per2PerES}.
Our \cs framework deals with this scenario in exactly the same way as the other $AB$ cases,
although the results are, as expected,
slightly worse than that of $2\%$ heterozygous carriers.

\subsection{The effect of noise}
\label{sec:noiseEffect}
Figures~\ref{fig:noiseEffectsOn1PerRegAndNoSampling},\ref{fig:noiseEffectsOn1PerRegAndErrors}
present the effects of three types of noise in the specific case of $f=1\%$.
In both figures the reference is the ``standard'' performance of $f=1\%$ which appeared
before in Fig.~\ref{fig:lanesAndNmax1Per2Per}, and includes sampling noise,
read error ($e_r=1\%$) and \ppt errors ($\sigma=0.05$).
We consider the case of sampling error separately from the other two sources, as its
impact is different.

\subsubsection{Sampling error}
\begin{figure}[thb!]
\begin{center}
\psfig{file=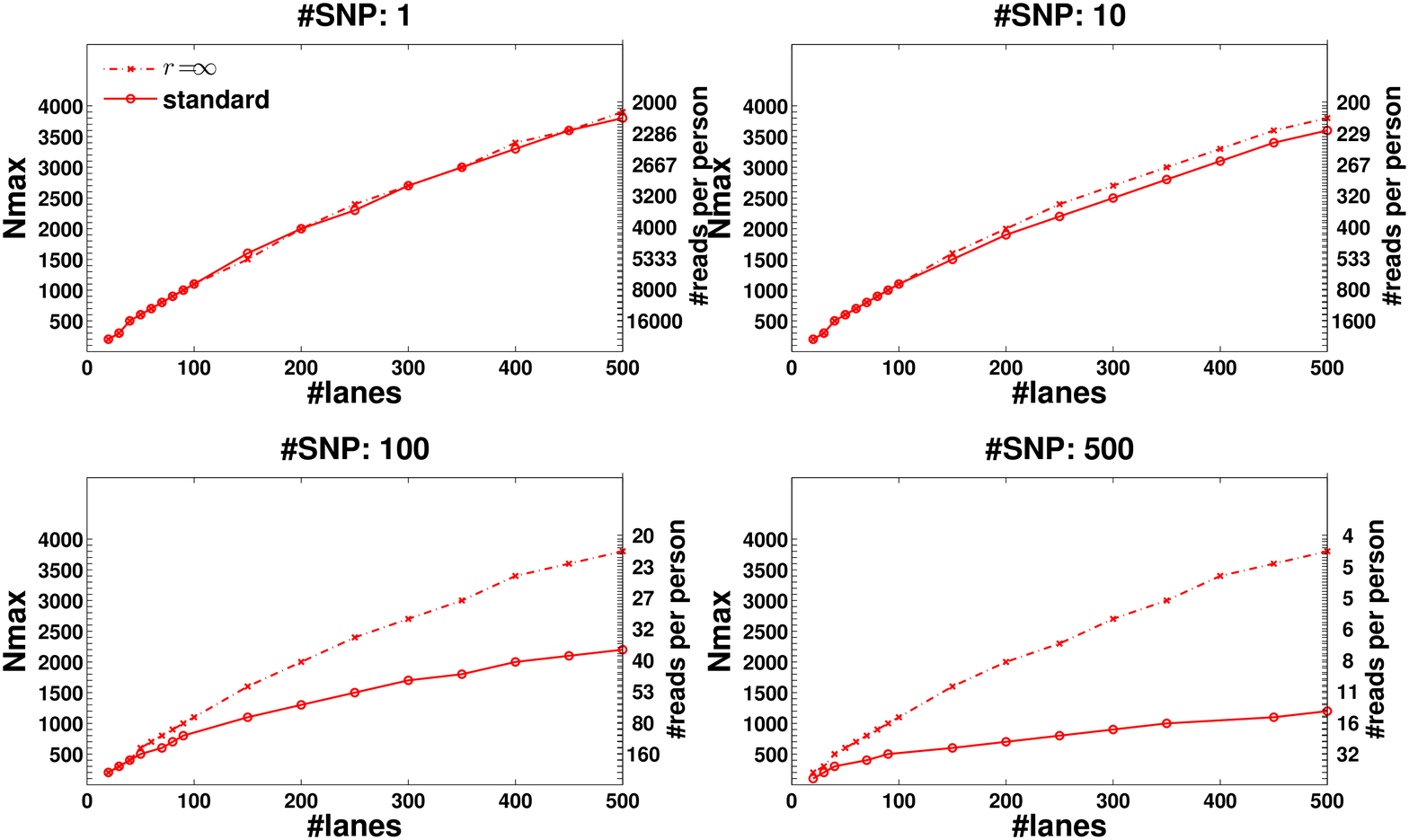,width=12cm}
\caption{Effect of sampling error.
The dashed line represent results obtained in the limit where the number
of reads goes to infinity thus sampling error is zero.
The solid line represents the realistic scenario
with the current number of reads used.
Sampling error is seen to be a significant factor when we treat many loci
together in the same lane ($L=100$ or more),
while for a few loci ($L=10$ or less) we already have enough coverage
to make sampling error negligible.}
\label{fig:noiseEffectsOn1PerRegAndNoSampling}
\end{center}
\end{figure}

Figure~\ref{fig:noiseEffectsOn1PerRegAndNoSampling} compares the ``standard'' performance
to the case where an infinite number of reads are available (although read error
and \ppt errors are still present.) Differences between the cases appear only when the number
of SNPs is high, thus the number of reads per person $c$ is insufficient.
In these cases $N_{max}$ is reduced by a factor of $2$ to $4$ with respect to
an infinite coverage. When the number of SNPs is small, and coverage is high,
we see no difference between the ``standard'' performance and the infinite read case.

\subsubsection{Read errors and \ppt errors}
\begin{figure}[thb!]
\begin{center}
\psfig{file=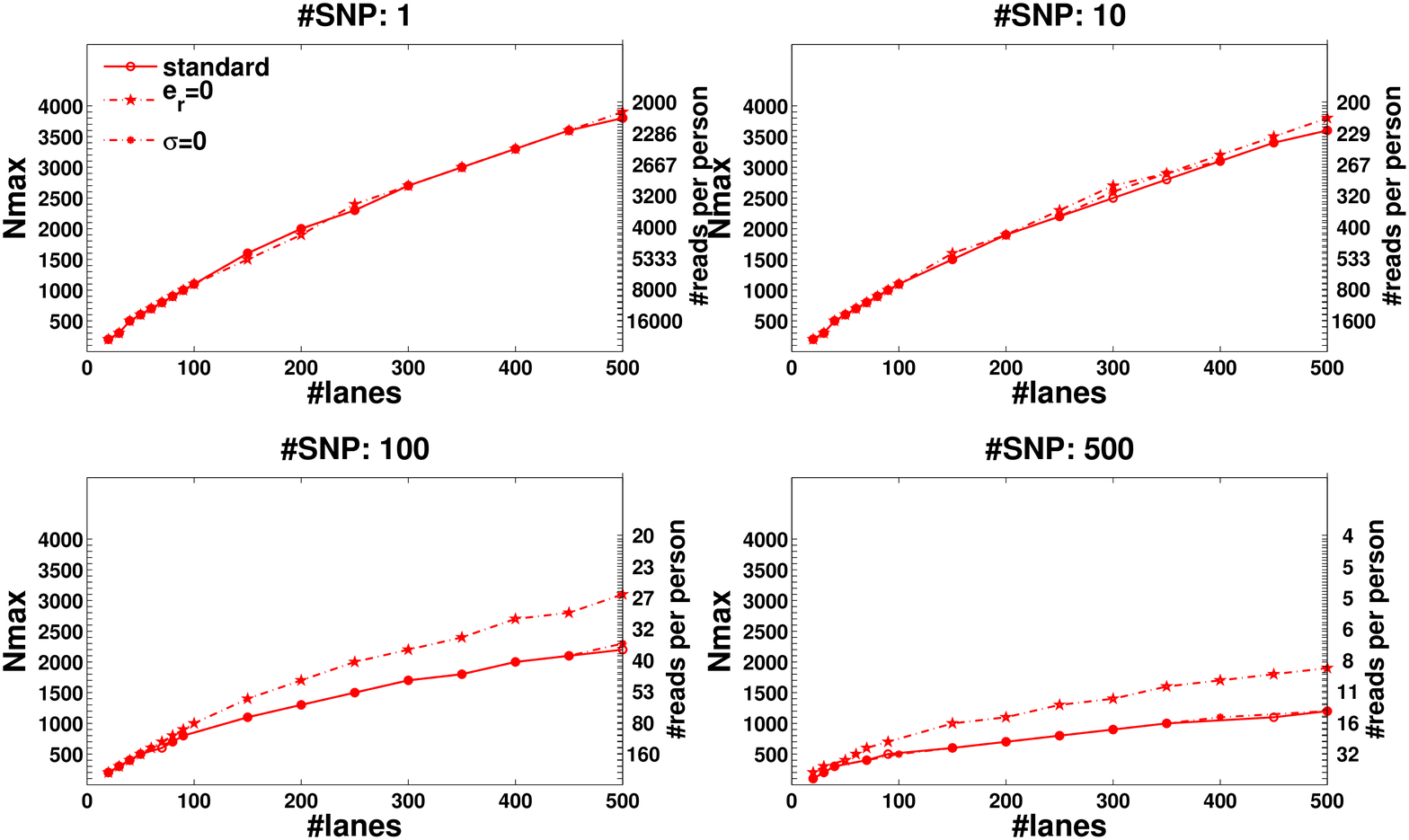,width=12cm}
\caption{Effect of read errors and \ppt errors. The two dashed lines represent results obtained
when assuming that reads are perfect and \ppt errors is zero, respectively.
The solid line represents a realistic scenario
with a read error of $1\%$ and \ppt errors of $\sigma = 0.05$.
While read error appears to have a significant
factor in reducing $N_{max}$, the effect of \ppt errors seems to be negligible.}
\label{fig:noiseEffectsOn1PerRegAndErrors}
\end{center}
\end{figure}

Figure~\ref{fig:noiseEffectsOn1PerRegAndErrors} compares the ``standard''
performance to two cases: one in which $e_{r}=0$ and another in which $\sigma=0$.
In the absence of read errors $N_{max}$ may be twice as large as when $e_{r}=1\%$.
Read errors take a significant effect on performance only when $L$ is large ($100$ or $500$),
since when coverage is high read errors are compensated for (see Eq.~(\ref{eq:CSopt_noise}).)
In all cases we have studied the results are quite robust to \ppt errors, thus noise introduced
by realistic pooling protocols should be easily overcome by the \cs reconstruction.

\subsection{Modifying the sensing matrix}
\label{sec:Msqrt}
In all simulations presented so far we have considered the case where each pool includes 
approximately $N/2$ individuals.
It may be desirable to minimize the number of individuals per pool \cite{Erlich:01},
as this can lead to a faster and cheaper preparation of each pool.
Here we shortly present the possibility of modifying $M$ into a {\it sparse} sensing matrix, thus accommodating
the requirement of having few individuals per pool.

\begin{figure}[thb!]
\begin{center}
\begin{tabular}{cc}
\psfig{file=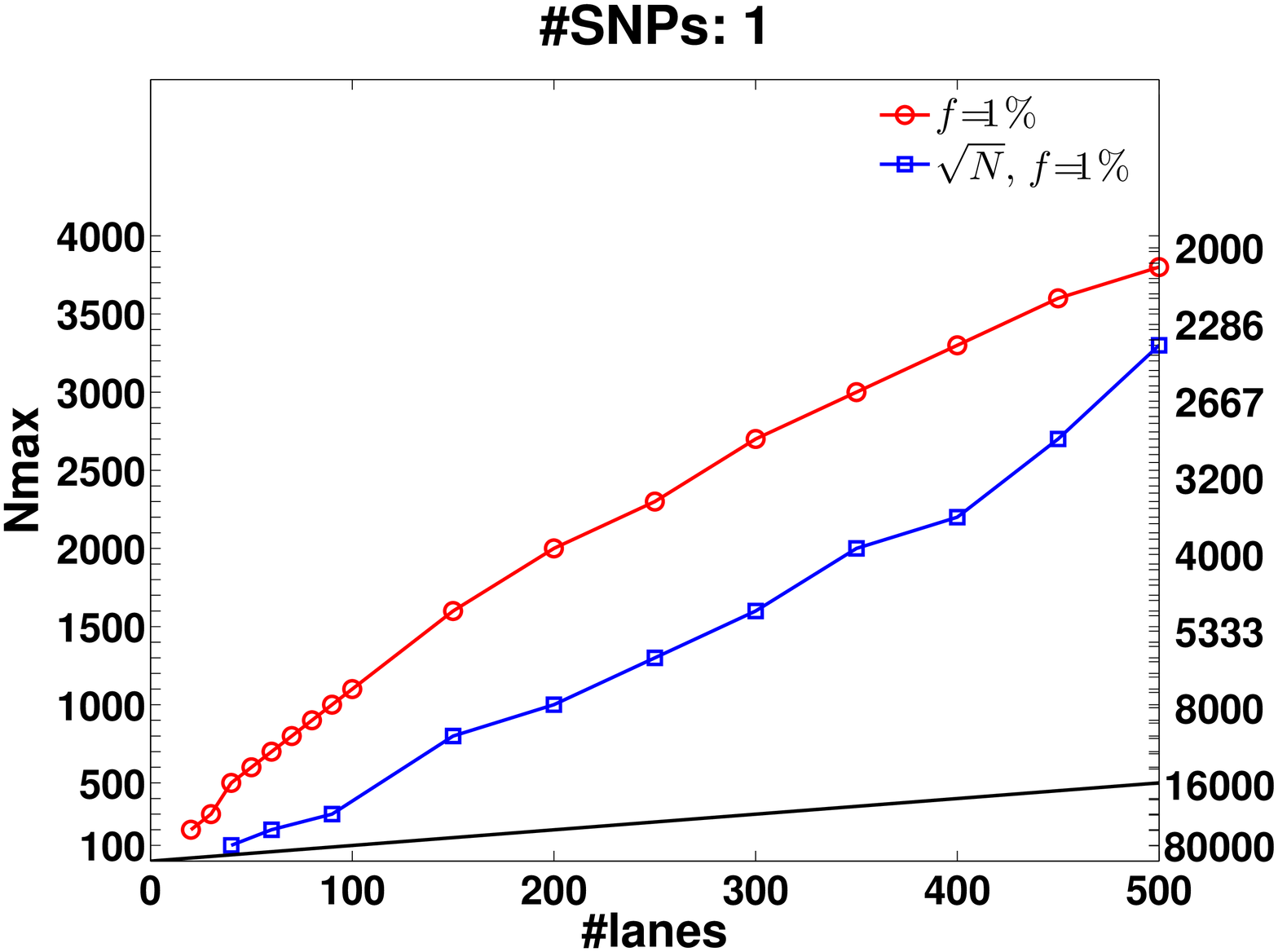,width=6cm} &
\psfig{file=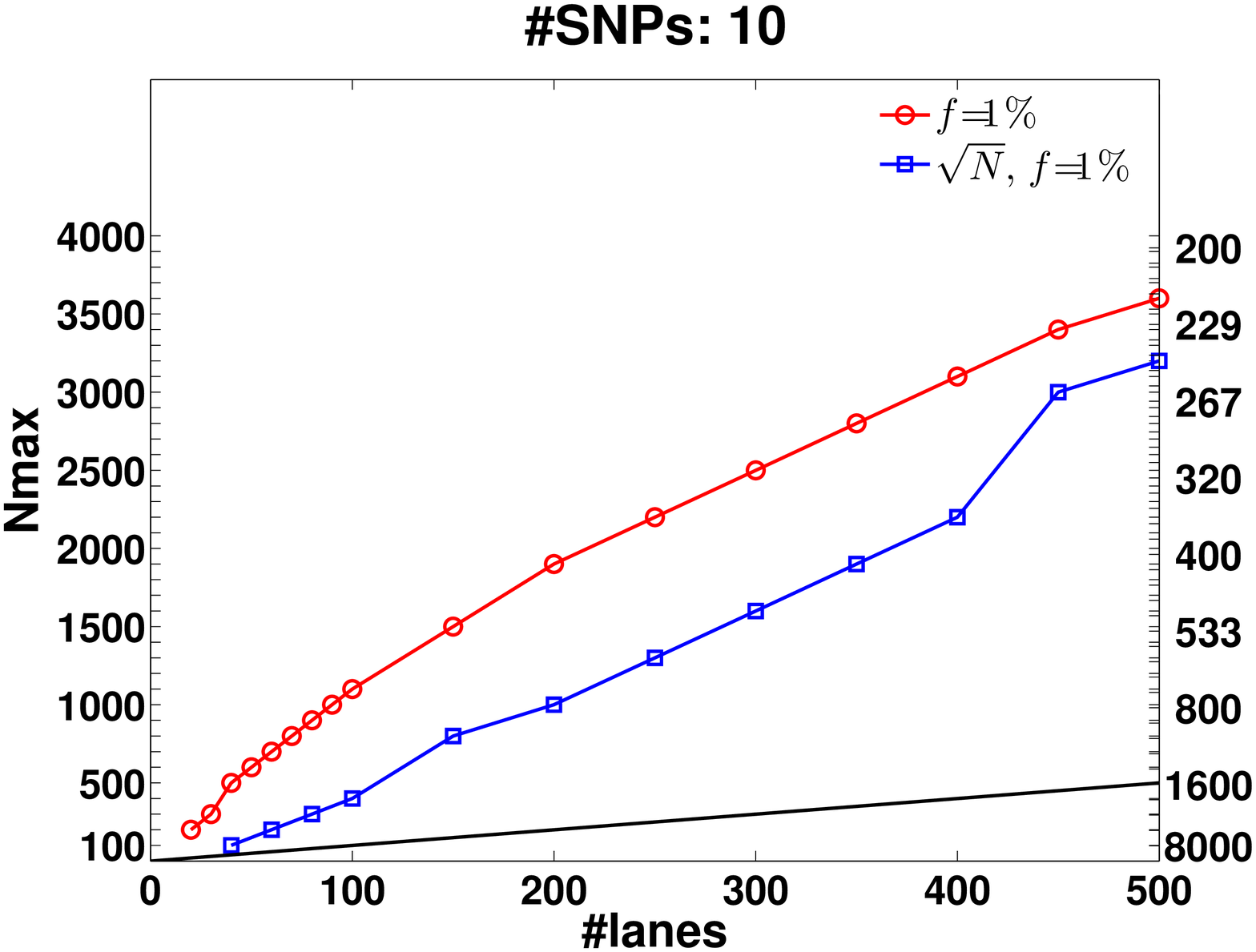,width=6cm}\\
\psfig{file=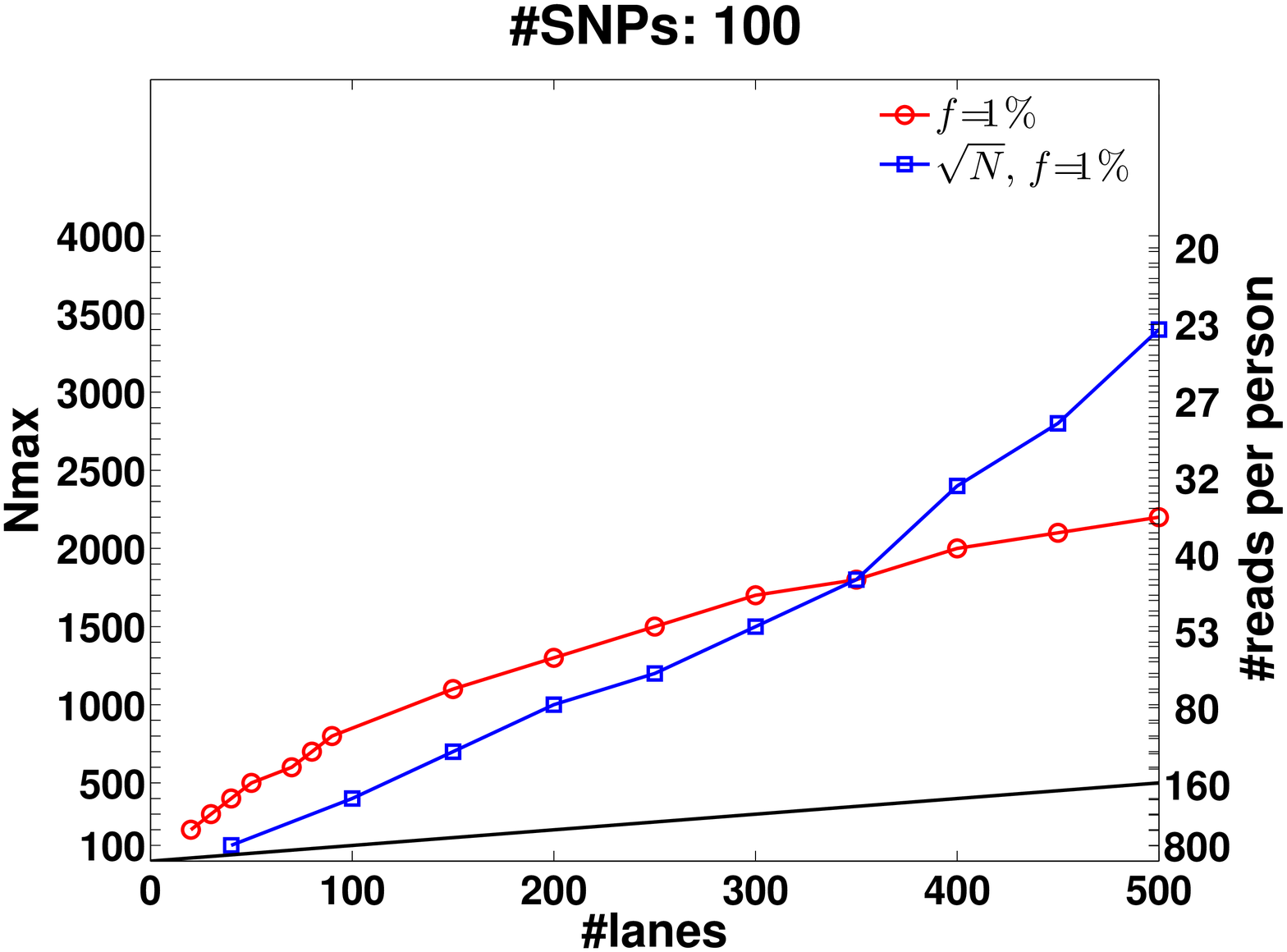,width=6cm} &
\psfig{file=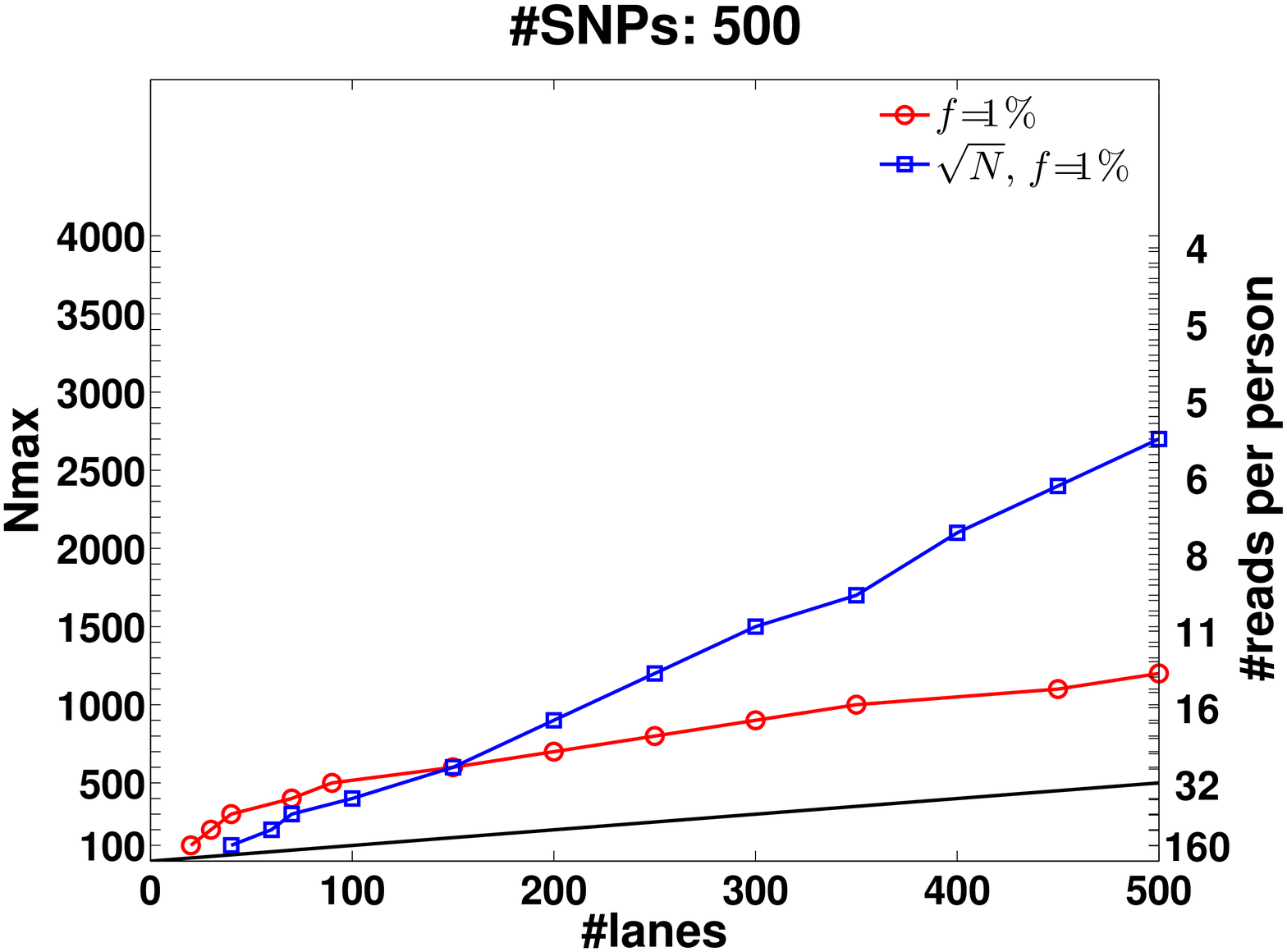,width=6cm}
\end{tabular}
\caption{The effect of applying pools of $\sqrt{N}$ or $N/2$ individuals (on average) for
the case of $f=1\%$.
Overall, results are comparable, yet each pooling design is preferable
for different settings of lanes and loci. The sparse ($\sqrt{N}$) design is more beneficial for
large number of lanes and longer target regions.
The average coverage   on the right axis of each panel corresponds to the $N/2$ case. 
The coverage in the $\sqrt{N}$ case is much larger since the total number of reads $R$ is 
divided among a smaller number of individuals.}
\label{fig:effectSqrt}
\end{center}
\end{figure}

Figure~\ref{fig:effectSqrt} presents the results of using only $\sqrt{N}$
individuals in each pool, for the case $f=1\%$ (marked as ``$\sqrt{N}, f=1$'').
For a small number of loci taken together ($L=1$ or $10$) the former dense Bernoulli(0.5) sensing
matrix achieves higher $N_{max}$ values.
However, when the number of loci is large ($L=100$ or $500$) and for large number of lanes,
it is preferable to use sparse pools of size $\sqrt{N}$.
The same qualitative behavior was observed for $f=0.1\%$ and $f=2\%$. The success of sparse
matrices in recovering the true genotypes is not surprising given theoretical and experimental
evidence \cite{berinde2008sparse}. Further research is needed in order to determine
the optimal sparsity of the sensing matrix for a given set of parameters.

\subsection{Combining barcodes and \cs}
\label{sec:resultsBarcodes}
\begin{figure}[thb!]
\begin{center}
\psfig{file=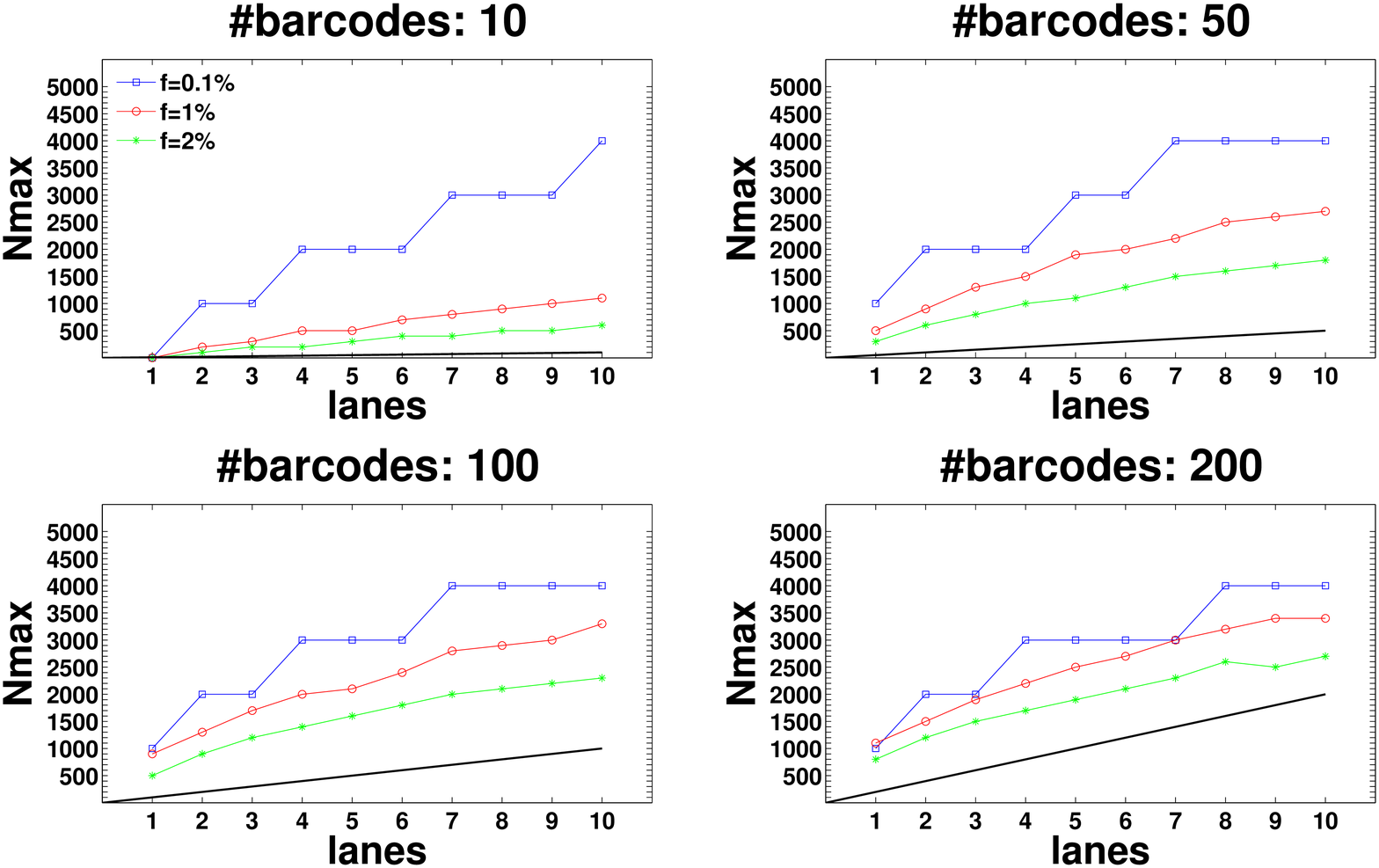,width=12cm}
\caption{Results obtained by combining the \cs approach with barcoding for $L=1$. 
Barcoding improves results by enabling a higher number of `effective' lanes
(although these lanes contain a smaller number of reads.)
The effect of adding a large number of barcodes is more pronounced for high minor-allele frequency.
For example, at $f=2\%$ and with $7$ lanes, we can treat roughly $300$ individuals with $10$ barcodes,
but around $2500$ individuals with $200$ barcodes.
The increase in power is sub-linear, as is seen by the fact that when we add more barcodes,
the performance becomes closer to that of the naive approach (shown in black) which increases
linearly with the number of lanes. Still, only at a very high number of barcodes the naive approach can
perform as well as the \cs design.}
\label{fig:barcodeAndNmax}
\end{center}
\end{figure}

Barcodes may also be combined with \cs so as to improve efficiency and further reduce the number
of required lanes.
The DNA in each pool (as opposed to the DNA of a specific individual) may be tagged using a
unique barcode (see Section~\ref{sec:MethodsBarcodes}.) Hence, in case we have $n_{bar}$ different
barcodes available, we apply $n_{bar}$ pools to a single lane,
with the price being that each pool contains only $R/n_{bar}$ reads.
Figure~\ref{fig:barcodeAndNmax} displays $N_{max}$ as a function of the number of lanes,
for different values of $n_{bar}$, and for different rare allele frequencies.

The black line in each figure represents the ``naive'' capacity,
which is simply $k\times n_{bar}$.
Incorporating even a small number of barcodes into our \cs framework
results in a dramatic increase in $N_{max}$. For the same problem parameters but without using barcodes,   
could not recover the minimal possible number of individuals $N_{max} = 1000$ for $f=0.1\%$ and could not 
reach more than $N_{max} = 100$ for $f=1\%,2\%$ (see Figs.~\ref{fig:lanesAndNmax1Pro},\ref{fig:lanesAndNmax1Per2Per}.)
Similarly to the non-barcodes case, the advantage over the naive approach is most prominent for $f=0.1\%$, 
but is still significant for $f=1\%,2\%$. As the number of barcodes increases, the difference in performance 
between different sparsities $f$ becomes smaller. As long as the coverage is kept high, it is still beneficial
to increase the number of barcodes, as it effectively increases linearly the number of lanes. At a certain
point, when many different barcodes are present in a single lane, coverage drops and sampling error becomes significant,
hence the advantage of adding more barcodes starts to diminish.

\section{Discussion}
\label{sec:discussion}
We have presented a method for identifying rare allele carriers via \cs\!\!-based group testing.
The method naturally deals with all possible scenarios of multiple carriers
and heterozygous or homozygous rare alleles.
Our results display the advantages of the approach over the naive one-individual-per-lane scenario:
it is particularly useful for the case of a large number of individuals and low frequencies of rare alleles.
We have also shown that our method can benefit from the addition of barcodes
for different pools~\cite{Erlich:01}, and still improve upon `standard' barcoding.

We view our main contribution as outlining a generic approach that puts together sequencing
and \cs for solving the problem of carrier identification. Following this mapping
we may apply the vast amount of \cs literature and benefit from any advancement
in this rapidly growing field.
We believe this is a major advantage over more `tailored' approaches,
e.g. \cite{Prabhu:01,Erlich:01},
although these methods may be superior to ours in specific cases.

Our method is {\it simple} in the sense that it applies \cs in the most
straightforward fashion.
We have used an off-the-shelf \cs solver and did not try to optimize any \cs
related parameter for the different scenarios - all parameters were kept fixed
for all types of simulations thus optimizing reconstruction algorithms is likely
to improve our results. Moreover, the method's performance as detailed in
Section~\ref{sec:Results} may be further improved, since our formulation of
the \cs problem has incorporated only part of the available information at hand.
Using additional information may reduce the number of lanes or
total reads needed to achieve a certain accuracy, as well as enable faster
algorithms for reconstruction, thus allowing us to deal with larger sample
sizes and larger regions.

As an example, the information that the input signal is trinary $(0,1,2)$
was considered only in the post-processing step after running GPSR,
although it may be incorporated into the optimization procedure itself.
Since most alleles are approximately distributed according to HW equilibrium,
the frequency of $2$'s, derived from the frequency of $1$'s is very low, and the input vector
is most often Boolean. Therefore, we could also modify the optimization so as to `punish'
for deviations from this pattern.
In addition, we have treated each rare allele independently,
although in case alleles originate from the same genomic region,
one could use linkage disequilibrium information and reconstruct haplotypes.
For example, if two individuals share a rare allele at position $i$,
they are also likely to share an allele, probably rare, at position $i+1$.
It would be a challenge to add these constraints and enable the reconstruction of several adjacent loci
simultaneously.
Another improvement may stem from more accurate modeling of specific errors
of sequencing machines, including quality scores which provide an estimate of error
probability of each sequenced base \cite{brockman2008quality,Kao:01}.
Such careful modeling typically results in far smaller error rates than the ones
we have conservatively used ($\sim 0.5\%-1\%$.)

Another possible improvement may be to apply adaptive group testing,
namely to decide whether to use lane $k$, based on reconstruction results of
lanes $1,\ldots, k-1$.
This enables an adaptation to unknown sparsity by simply
generating pools one by one, each time solving the \cs problem and checking if we
get a sparse and robust solution.
Once the solution stabilizes, we can stop our experiments
thus not `wasting' unnecessary lanes (it may also
be beneficial to change the pooling design of the next measurement and
allow deviations from the randomized construction we have shown, once statistical
information regarding the likely carriers is starting to accumulate).

The drawbacks of our method stem from the limitations of \cs and sequencing technology.
First of all, in case rare allele frequency is high enough, the sparsity assumption at the heart
of the \cs theory breaks down, and it may be problematic for \cs
to reconstruct the signal. The highest frequency possible for \cs to perform well in this application
was not determined, but one should expect a certain frequency above which it is
no longer beneficial to apply \cs and the naive one-individual-per-lane approach is preferable.
The simulations we have performed estimate this frequency to be over $5\%$ in
most cases, thus taking effect only for the case of {\it common} alleles.

The number of individuals pooled together and the length of the targeted regions are limited
by the total capacity of a given lane. We have shown that efficient reconstruction is possible when a
significant number of individuals is pooled together, and when one can target efficiently a rather small
genomic region. However, efficiency could be further improved if we could
maintain a large enough coverage while further increasing the pool and region size.
We expect that in the foreseeable future sequencing technology would yield higher number of reads per lane,
which would allow larger pool size and genomic regions to be treated by our approach.

Another difficulty in our approach is related to the issue of randomness of the {\it sensing matrix} $M$.
This randomness may be discarded by simply fixing a certain instance of the sensing matrix,
although randomness in this case may be viewed as an advantage of \cs - almost any (random) matrix
would enable reconstruction, as opposed to intricate pooling schemes which need to be carefully designed.

The last drawback we should mention is related to the fact
that each pool contains approximately half the individuals in the group.
This may be problematic in cases where pooling preparation might be slow and costly,
and we need to minimize the number of individuals in each pool \cite{Erlich:01}.
In this case it may be interesting to apply a sparser pooling design.
As shown in Section~\ref{sec:Msqrt}
there are scenarios in which it is advantageous to assign only $\sqrt{N}$
individuals to a pool. Therefore, one needs to optimize the pool design
together with other parameters, e.g. number of loci and lanes considered.
This issue remains for future study.

Another major direction we intend to pursue is to test the \cs approach experimentally.
To the best of our knowledge, no available data
is completely suitable for our purposes, thus we can not `adapt'
current data sets and `simulate' such an experiment.
Our approach is most beneficial for relatively high coverage and a large number of individuals,
thus designing a suitable experiment is needed.

Finally, while we have demonstrated the benefits of \cs\!-based group testing approach for
genotyping, any genetic or epigenetic variant is amenable to our approach, as long as
it can be detected by next generation sequencing technology and is rare in the population of interest.
For example, Copy Number Variations (CNVs), important for studying both normal population
variations and alterations occurring in cancerous tissues, provide a natural extension to our framework.
In this case the number of reads serves as a proxy to the copy number at a given locus,
and the vector to be reconstructed contains the (integer) copy number levels of each individual,
rather than their genotypes.
Another example is given by rare translocations, often present in various tumor types - where
an evidence for a translocation may be provided by a read whose head is mapped to one genomic region
and whose tail is mapped to another distal region (or by two paired-end reads, each originating from a different genomic region.)
Carriers of a particular rare translocation may be discovered using this approach. The extension of our method to these and
perhaps other novel applications provides an exciting research direction we plan to pursue in the future.

\bibliography{./bib_metagenomics}

\appendix
\section{Coping with unknown read error}
\label{app:unknown_p_err}
We assume that the read error $e_r$ is unknown to the researcher, but is constant across all
lanes. One can introduce a slight modification to our procedure, which enables the learning of
$e_r$ from our pooling data. We replace $\textbf{z}$ in Eq.~(\ref{eq:CSopt}) by the convolution:
\be
\textbf{z} * e_r \equiv \textbf{z} + e_r - 2 \textbf{z} e_r
\ee

The additive factor $e_r - 2 \textbf{z} e_r$ is different for different values of $\textbf{z}$,
but its dominant part is $e_r$. We can approximate it by $x_{N+1} \equiv e_r - 2 \bar{z} e_r$,
obtained from averaging the term $2 \textbf{z} e_r$ over all $\textbf{z}$ values (i.e. $\bar{z}$ is the
mean value of the vector $\textbf{z}$). We can now reformulate
the \cs problem by adding one extra variable. Specifically, the unknown vector $\textbf{x}$
is replaced by $\textbf{x'} = (\textbf{x}, x_{N+1})$ and Eq.~(\ref{eq:CSopt}) is replaced by:
\be
{\textbf{x'}}^* = \underset{\textbf{x'}}{argmin} ||\textbf{x'}||_1 \quad s.t. \: \:
||\frac{1}{2} \hat{M'}\textbf{x'}-\frac{1}{r}\textbf{z'}||_2 \leq \eps
\label{eq:cs_problem_L1_unknown_p_err}
\ee

where $M'$ is built from $M$ by adding a constant column to its right as its $N+1$'s column with all its values set to $-1$.

\end{document}